\documentclass[12pt,aps,floats,showpacs,amssymb,tightenlines]{revtex4}

\usepackage{color}
\usepackage{graphicx}
\usepackage{amsmath}
\usepackage{amsfonts}
\usepackage{amscd}
\usepackage{epsfig}
\usepackage{amssymb}
\usepackage{tabularx}
\usepackage{epstopdf}

\begin{document}

\title{Perturbative evolution of the static configurations, quasinormal modes and
quasi normal ringing in the Apostolatos - Thorne cylindrical shell model.}
\author{Reinaldo J. Gleiser} \email{gleiser@fis.uncor.edu} \author{Marcos A. Ramirez}

\affiliation{Instituto de F\'{\i}sica Enrique Gaviola and FAMAF,
Universidad Nacional de C\'ordoba, Ciudad Universitaria, (5000)
C\'ordoba, Argentina}

\begin{abstract}
We study the perturbative evolution of the static configurations, quasinormal modes and
quasi normal ringing in the Apostolatos - Thorne cylindrical shell model. We consider first an expansion in harmonic modes and show that it provides a complete solution for the characteristic value problem for the finite perturbations of a static configuration. As a consequence of this completeness we obtain a proof of the stability of static solutions under this type of perturbations. The explicit expression for the mode expansion are then used to obtain numerical values for some of the quasi normal mode complex frequencies. Some examples involving the numerical evaluation of the integral mode expansions are described and analyzed, and the quasi normal ringing displayed by the solutions is found to be in agreement with quasi normal modes found previously. Going back to the full relativistic equations of motion we find their general linear form by expanding to first order about a static solution. We then show that the resulting set of coupled ordinary and partial differential equations for the dynamical variables of the system can be used to set an initial plus boundary values problem, and prove that there is an associated positive definite constant of the motion that puts absolute bounds on the dynamic variables of the system, establishing the stability of the motion of the shell under arbitrary, finite perturbations. We also show that the problem can be solved numerically, and provide some explicit examples that display the complete agreement between the purely numerical evolution and that obtained using the mode expansion, in particular regarding the quasi normal ringing that results in the evolution of the system. We also discuss the relation of the present work to some recent results on the same model that have appeared in the literature.

\end{abstract}


\maketitle
\section{Introduction}

The general analysis of the evolution of material systems interacting with their own gravitational field is undoubtedly a very important and very difficult problem in General Relativity. As usual in any physical theory, one can gain insight into the more difficult issues by considering idealized simplified models that, nevertheless, still preserve some of the relevant features of the general problem. One of these simplified examples was described some time ago by Apostolatos and Thorne \cite{apostol}. The material contents in this example is given by a cylindrical shell of counter rotating particles, and one is interested in the dynamics that results as a consequence of its interaction with its own gravitational field.  As shown in \cite{apostol}, once some appropriate choice of coordinates is made, it is a straightforward matter to obtain a set of coupled equations for the evolution of the matter and field variables, which is described in some detail in the next Section. In their original paper Apostolatos and Thorne were interested in general features of the evolution, in particular the possibility of naked singularities forming as the result of the collapse of these structures, but the detailed evolution of the shell in a situation close to one of its static configuration was only qualitatively mentioned, with no detailed results. This problem is interesting in particular because in the Newtonian approximation the shell is either in a static configuration or performs periodic oscillations about this static configuration, but, once the radiative modes of the gravitational field are introduced, one expects these oscillations to be eventually damped as the results of the transfer of energy to the radiative modes. Thus, in the fully relativistic description one expects to find the characteristic ``Quasi Normal Ringing'' (QNR), related to the ``Quasi Normal Modes'' (QNM)\cite{nollert} of the system.

A recent analysis of the dynamics of the Apostolatos and Thorne model was given by Hamity, et.al. \cite{hamity}. We notice, however, that the approximations introduced in the explicit examples considered in \cite{hamity} are such that the radiative modes are effectively neglected, and, therefore, the results obtained are possibly relevant only close to the Newtonian limit. A detailed analysis of the full relativistic equations in the linearized approximation was carried out by the present authors in \cite{gleram}, where we obtained the general solution for the harmonic modes with real frequencies, and considered the solutions obtained by general linear combinations of these modes. Although in all the examples considered in \cite{gleram} we found a stable evolution under perturbations of (essentially) compact support, at least two questions remained unanswered. The first was the possible completeness of the mode expansion, and its relation to the initial value problem for the system. The second was the apparent lack of QNR in the example solutions. The main purpose of the present work is to give answers to those questions. We first review in Sections 2 and 3, mostly for completeness, the main features of the model. Then, in Section 4, we review the derivation of the periodic solutions in the linearized regime, introducing a modified form, as compared with that given in \cite{gleram}, that allows us to establish, in Section 5, the completeness of expansion in the context of the characteristic value problem for the system, and, therefore, the completeness of the mode expansion. This completeness, in turn, as discussed in Section 6, implies the linear stability of static configurations under this type of perturbations, a result that differs from some conclusions reached in \cite{hamity}, and recent results obtained by Kurita and Nakao \cite{nakao}. Next, in Section 7, we discuss the existence of QNMs for the system. As shown there, these are related to the zeros of a complex expression that appears in the relation between the incoming and outgoing wave amplitudes as functions of the frequency. This expression has a complicated form involving Bessel functions that makes it very difficult its analysis. For this reason we have not been able to obtain explicit expressions for the frequencies of the QNM, or even to ascertain whether there is finite number or infinite number of these frequencies. Nevertheless, by resorting to numerical methods we found at least a couple of QNM corresponding to a range of parameters characterizing static configurations. The values obtained are also displayed in Section 7. In Section 8 we consider the numerical evaluation of the integrals obtained in Section 7 that provide the evolution of the system for arbitrary characteristic data. The examples chosen display clearly the quasi normal ringing associated to one of the QNM. The reason why this QNR was not seen in \cite{gleram} is discussed in Section 12, using the Price-Husain toy model as an illustration.

In Section 9 we consider again the full relativistic equations and assuming a general linear departure from a static configuration obtain the general linearized equations for the system. Then, in Section 10 we prove the existence of an associated positive definite constant of the motion that puts absolute bounds on the dynamic variables of the system, establishing the stability of the motion of the shell under arbitrary, finite perturbations. In Section 11 we show that the corresponding set of coupled ordinary and partial differential equations for the relevant dynamical variables can be solved numerically as an initial plus boundary values problem, without resorting to expansions in terms of Bessel or other functions. We provide a couple of examples, using the same shell parameters as in the mode expansion, but a different form for the incoming wave form. The results, that clearly display quasi normal ringing are in complete agreement with those of the mode expansion, and with the computed QNM frequencies, although the techniques used in both approaches are totally different, confirming their separate  correctness.

We end the paper with some comments and conclusions, as well as comments on related work by other authors, in particular to that of reference \cite{nakao}.

\section{The Apostolatos - Thorne model}

The Apostolatos - Thorne model \cite{apostol} describes the dynamics
of a self gravitating cylindrical shell of counter rotating
particles. Both the inner ($M^-$) and  outer ($M^+$) regions of the shell are vacuum
space times with a common boundary $\Sigma$. The corresponding metrics may be written in the form,
\begin{equation}\label{ATeq1}
ds^2_{\pm} = e^{2\gamma_{\pm}-2 \psi_{\pm}}
\left(dr^2-dt^2_{\pm}\right)+ e^{2 \psi_{\pm}}dz^2 +e^{-2
\psi_{\pm}}r^2 d\phi^2
\end{equation}
where the $(+)$ sign corresponds to the outside and $(-)$ to the
inner regions. The functions $\psi$, and $\gamma$ depend only on $r,t$
and satisfy the equations:
\begin{equation}\label{ATeq2}
    \psi_{,rr} +\frac{1}{r}\psi_{,r} -\psi_{,tt} = 0
\end{equation}
\begin{equation}\label{ATeq3}
\gamma_{,t} = 2 r \psi_{,r} \psi_{,t}\;\;\; , \;\;\; \gamma_{,r} = r
\left[(\psi_{,r})^2+(\psi_{,t})^2\right]
\end{equation}
The shell is located on the hypersurface $\Sigma$ given by $r=R(\tau)$,
where $\tau$ is the proper time of an observer at rest on the shell.
We may interpret $\psi(r, t)$ as playing the role of a gravitational
field whose static part is the analogue of the Newtonian potential.
The time dependent solutions of (\ref{ATeq2}) represent
gravitational waves (Einstein-Rosen). Equation (\ref{ATeq2}) is the
integrability condition of Eqs. (\ref{ATeq3}). The coordinates $(z,
\phi, r )$ and the metric function $\psi$ are continuous across the
shell $\Sigma$, while $t$ and the metric function $\gamma$ are
discontinuous. Smoothness of the spacetime geometry on the axis $r =
0$ requires that $ \gamma = 0$, $\psi$ finite at $r = 0$, and $\partial \psi/\partial r|_{r=0} =0$. The
junction conditions of $M^-$ and $M^+$ through $\Sigma$ require the
continuity of the metric and specify the jump of the extrinsic
curvature $K^{\pm}$ compatible with the stress energy tensor on the
shell. The induced metric on $\Sigma$ is given by
 \begin{equation}\label{ATeq4}
ds^2_{\Sigma} =-d \tau^2 + e^{2 \psi_{\Sigma}} dz^2 +  e^{-2
\psi_{\Sigma}} R^2 d\phi^2
 \end{equation}

Here $\psi_{\Sigma}(\tau) =
\psi_+(R(\tau),t_+(\tau))=\psi_-(R(\tau),t_-(\tau))$. The evolution
of the shell is characterized by $R(\tau)$, which is the radial
coordinate $r$ at the shell's location and $\tau$ the proper time of
an observer at rest on $\Sigma$. If we assume that the shells is
made up of equal mass counter rotating particles, the Einstein field
equations on the shell may be put in the form,
\begin{equation}\label{ATeq5}
\psi^+_{,n}-\psi^-_{,n} = -\frac{2 \lambda} {  \sqrt{R^2+
e^{2\psi_{\Sigma}} J^2}  }
\end{equation}
\begin{equation}\label{ATeq6}
X^+-X^- = -\frac{4 \lambda \sqrt{R^2+e^{2 \psi_{\Sigma}}J^2}}{R}
\end{equation}
where the constants $\lambda$ and $J$ are, respectively, the proper
mass per unit Killing length of the cylinder and the angular
momentum per unit mass of the particles. The other quantities in
(\ref{ATeq5},\ref{ATeq6}) are given by,
\begin{equation}\label{ATeq7}
X^{\pm} \equiv \frac{\partial t_{\pm}}{\partial \tau} =
+\sqrt{e^{-2(\gamma_{\pm}-\psi_{\Sigma})} +\dot{R}^2}
\end{equation}
\begin{equation}\label{ATeq8}
\psi^{\pm}_{,n} = \psi^{\pm}_{,r} X^{\pm} +\psi^{\pm}_{,t} \dot{R}
\end{equation}
where a dot indicates a $\tau$ derivative, and we also have,
\begin{eqnarray}
\label{ATeq9} \frac{d^2 R}{d\tau^2} &=&  \dot{R}\dot{\psi_{\Sigma}}
- R
\left[(\dot{\psi_{\Sigma}})^2 +(\psi^-_{,n})^2\right]  \nonumber \\
& & +\frac{ R^2 \psi^-_{,n} X^-}{R^2+e^{2 \psi_{\Sigma}}J^2}
-\frac{\lambda R^2 X^-}{(R^2+e^{2 \psi_{\Sigma}}J^2)^{3/2}}
+\frac{J^2 e^{2 \psi_{\Sigma}}X^-X^+}{R(R^2+e^{2 \psi_{\Sigma}}J^2)}
\end{eqnarray}

These equations together with (\ref{ATeq2},\ref{ATeq3}) determine
the evolution of the shell and of the gravitational field to which
it is coupled.

\section{Static solutions}

We briefly review the conditions for a static solutions, corresponding to a shell of constant radius $R(\tau)=R_0$. We restrict to the case of a regular interior. In this case the interior must be flat and we may take
$\gamma_-=0$, $\psi_- =0$, implying $\psi_{\Sigma}=0$. For the
exterior field the general static solution (satisfying $\psi_{\Sigma}=0$) is of the form,
\begin{equation}\label{eq1a}
\psi^+(r)= - \kappa \ln(r/R_0) \;\;\; , \;\;\; \gamma^+(r)= \gamma_0
+\kappa^2 \ln(r/R_0)
\end{equation}
Then,
\begin{equation}\label{eq2a}
X^+(r)= e^{-\gamma_0} \;\;\; , \;\;\; X^-=1
\end{equation}
Since $\dot{R}=0$, $\dot{\psi}_{\Sigma}=0$, and $\psi^-_{,n}=0$, we
find,
\begin{equation}\label{eq3a}
\kappa = 2 \frac{J^2}{R_0^2} \;\;\;,\;\;\; \gamma_0 =2
\ln\left[(R_0^2+2J^2)/R_0^2 \right]
\end{equation}
and,
\begin{equation}\label{eq4a}
\lambda = \frac{ J^2 R_0 \sqrt{J^2+R_0^2}}{(2 J^2+R_0^2)^2}
\end{equation}
It is important to notice that this relation can be satisfied for real $R_0$, $J$ and $\lambda$ only if $\lambda \leq \lambda_c$, with
\begin{equation}\label{eq4aa}
\lambda_c = {\frac {\sqrt {3+\sqrt {17}} \left( \sqrt {17}-1
 \right) }{4(9+\sqrt {17})}} = 0.15879...
\end{equation}
Therefore, the system admits static solutions only if $\lambda \leq \lambda_c$ .

It will be convenient to write $J=x R_0$. Replacing in (\ref{eq4a}), we find,
\begin{equation}\label{eq5a}
\lambda = \frac{ x^2 \sqrt{1+x^2}}{(2 x^2+1)^2}
\end{equation}
which we now consider as an equation for $x$ as a function of $\lambda$.

It is now easy to check that there are no real solutions for $x$ if $\lambda > \lambda_c$. For $\lambda = \lambda_c$ we have a double solution $x = x_c = (\sqrt {\sqrt {17}-1})/2=0.88361\dots$. The interesting region for our analysis is the range $0 <\lambda < \lambda_c$ where we have {\em two} real and positive solutions for $x$, one larger and one smaller than $x_c$, which approach respectively $+\infty$ and $0$ as $\lambda \to 0$. This implies that for appropriate $\lambda$ and $J$ we have {\em two} static solutions. It was concluded in previous analysis \cite{hamity} that, at least from a perturbative point of view, one of these solutions (the one with the larger $R_0$) would be stable and the other (with the smaller $R_0$) would be unstable. This conclusion, as shown in the following Sections, is {\em not} supported by the present analysis of the perturbative evolution that follows from initial data in the neighborhood of either static configuration. In the next Section we derive the appropriate equations for a perturbative analysis, and then show how this construction can be used to solve the evolution equations given characteristic data.

\section{Linearized periodic solutions with a regular interior}

In this Section we consider the problem of finding linearized periodic solutions for our system imposing the condition of a regular interior. This problem was already considered in \cite{gleram}, where the main idea was to obtain stationary periodic solutions, but left open the problem of relating these solutions to the initial value problem. Here we will consider a slightly modified approach, where we express the solutions in $M_+$ in terms of incoming and outgoing wave solutions for the radiative part of $\psi_+$.

We first notice that if we impose regularity on the symmetry axis $r=0$, and assume that the interior region
is empty but may contain gravitational radiation, then $\psi_-$ admits periodic solutions of the form \cite{absomega},
\begin{equation}\label{eq1d}
\psi_-(t_-,r)=  A_1 J_0(|\Omega_-| r)  e^{i \Omega_- t_-}
\end{equation}
where $J_0$ is a Bessel function, and we consider $A_1$ as a quantity of first order in perturbation theory. Then, restricting again to linearized order we may set,
\begin{equation}\label{eq2d}
\gamma_-(t_-,r)= 0
\end{equation}
since, in this case, from (\ref{ATeq3}), $\gamma_-$ is of second order, and, therefore, also to the appropriate order, we may also set,
\begin{equation}\label{eq3d}
 t_-(\tau)= \tau
\end{equation}
We assume a periodic perturbation around an equilibrium
configuration characterized by $R_0$ and $J$ and therefore take,
\begin{equation}\label{eq4d}
 R(\tau)= R_0+ \xi_0 e^{i \Omega \tau}
\end{equation}
where we assume that $\xi_0$ is also of first order.

We need to specify now a corresponding solution for $\psi_+$. In our previous analysis, \cite{gleram}, we noticed that the general periodic perturbation for $\psi_+$ may be written as,
\begin{equation}\label{eq4da}
\psi_+(t_+,r) = -\kappa \ln\left(\frac{r}{R_0}\right)+\left[A_2 J_0(|\Omega_+| r) + B_2 Y_0(|\Omega_+| r)\right] e^{i \Omega_+ t_+}
\end{equation}
where $Y_0$ is a Bessel function and $A_2$ and $B_2$ are first order quantities. To this order we have,
\begin{equation}\label{eq10da}
\gamma_+(t_+,r)= \gamma_0 + \kappa^2
\ln\left(\frac{r}{R_0}\right) - 2 \kappa \left[A_2 J_0(|\Omega_+| r) + B_2 Y_0(|\Omega_+| r)\right] e^{i \Omega_+ t_+}
\end{equation}
where $\kappa$ and $\gamma_0$ are given by (\ref{eq3a}).
A straightforward calculation \cite{gleram}
shows that consistency of the equations at first order requires
$\Omega_-=\Omega$,
\begin{equation}\label{eq11d}
\Omega_+ = \frac  {(R_0^2+2 J^2)^2}{R_0^4} \Omega =e^{\gamma_0}
\Omega
\end{equation}
and,
\begin{equation}\label{eq12d}
t_+ = \frac {R_0^4}  {(R_0^2+2 J^2)^2}  \tau =e^{-\gamma_0} \tau
\end{equation}
Replacing now in (\ref{ATeq5}), (\ref{ATeq6}), and (\ref{ATeq9}), and
expanding to first order, we find a set of three linear independent
equations for $A_1$, $A_2$, $B_2$, and $\xi_0$, for every choice of $\Omega$, $R_0$ and $J$, and, therefore, three of these quantities can be solved in terms of the fourth. Since all the evolution equations are linear, we may consider linear superpositions of these solutions to obtain more general (non periodic) solution. In \cite{gleram}, for simplicity, we chose $\xi_0$ as the independent variable, and obtained corresponding expressions for the other three amplitudes. Assuming an arbitrary dependence of $\xi_0$ on $\Omega$, given by the function $\xi(\Omega)$,  $R(\tau)$ is given by,
\begin{equation}\label{eq12da}
R(\tau) = R_0 +\int_{-\infty}^{+\infty} \xi(\Omega) \exp(i \Omega \tau) d \Omega
\end{equation}
while for $\psi_-(r,t_-)$ and $\psi_+(r,t_+)$ we obtain expression in the form of Fourier transforms of $\xi(\Omega)$ multiplied by complicated expressions involving Bessel functions of $r \Omega_{\pm}$ and $R_0 \Omega_{\pm}$. Details are given in \cite{gleram}. We notice that although in principle, (\ref{eq12da}) is completely general, and one has a definite $\xi(\Omega)$ for any possible $R(\tau)$, in practice, what one would like to solve is the problem of the evolution of the system given some initial condition, and because of the structure of the equations that resulted in \cite{gleram}, the best one could do was to assume some form for $\xi(\Omega)$, and find the corresponding evolution of the system. Carrying out this program it was found there that there are rather general evolutions, in the sense that they represent the interaction of the shell with an incoming gravitational pulse of rather arbitrary shape, for which the response of the shell is given precisely by $\xi(\Omega)$. An important question here is actually how general these solutions of the problem are, namely, whether they include every possible evolution of the system, or on the contrary, they only represent a restricted set. Since we had no proof of the completeness of the set of modes, this question is highly non trivial. Further, the scheme developed in  \cite{gleram} did no appear to show the quasi normal modes and quasi normal ringing expected in a system such as this, that has a simple Newtonian limit, where the shell can execute periodic motions. One would expect that the presence of radiative modes in the general relativistic version should lead to damped oscillations, at least in some appropriate limit, and this was not immediately apparent in the discussion given in \cite{gleram}. As we shall show here, both the completeness problem and the question of the QNR can be answered by a simple change in the formalism presented in \cite{gleram}.
This can be achieved as follows. The expression,
\begin{equation}\label{eq5d}
\psi_{in}(t_+,r)=F\; e^{i \Omega_2 t_+}\left[J_0(|\Omega_2|r)+i \frac{\Omega_2}{|\Omega_2|}Y_0(|\Omega_2|r)\right]
\end{equation}
where $F$ and $\Omega_2$ are constants, is a solution of (\ref{ATeq2}) for $\psi_+$. For large $r$, using the asymptotic expansions for the Bessel functions $J_0$ and $Y_0$, we find,
\begin{equation}\label{eq6d}
\psi_{in}(t_+,r) \simeq F \sqrt{\frac{1}{\pi r |\Omega_2|}}\left(1- i \frac{\Omega_2}{|\Omega_2|}\right)  e^{i \Omega_2 (t_++r)}
\end{equation}
and, therefore, (\ref{eq5d}) represents a purely incoming wave solution of (\ref{ATeq2}).

Similarly,
\begin{equation}\label{eq7d}
\psi_{out}(t_+,r)=G\; e^{i \Omega_2 t_+}\left[J_0(|\Omega_2|r)-i\frac{\Omega_2}{|\Omega_2|}Y_0(|\Omega_2|r)\right]
\end{equation}
where $G$ is a constant, is a solution of (\ref{ATeq2}), for $\psi_+$, which, for large $r$ has the asymptotic behaviour,
\begin{equation}\label{eq8d}
\psi_{out}(t_+,r) \simeq G \sqrt{\frac{1}{\pi r |\Omega_2|}}\left(1+ i \frac{\Omega_2}{|\Omega_2|}\right)  e^{i \Omega_2 (t_+-r)}
\end{equation}
and, therefore, (\ref{eq7d}) represents a purely outgoing wave solution of (\ref{ATeq2}). We notice now that we have,
\begin{equation}
\label{eq8da}
  \psi_{in}(t_+,r)+\psi_{out}(t_+,r) =  \left[(F+G) J_0(|\Omega_2|r) + i \frac{\Omega_2}{|\Omega_2|} (F-G)Y_0(|\Omega_2|r)\right] e^{i \Omega_2 t_+}
\end{equation}
Therefore, instead of (\ref{eq4da}) we may write the full solution for $\psi_+(t_+,r)$ in the form,
\begin{equation}\label{eq9d}
\psi_+(t_+,r)= -\kappa \ln\left(\frac{r}{R_0}\right)+\psi_{in}(t_+,r)+\psi_{out}(t_+,r)
\end{equation}
where we have the identifications:
\begin{eqnarray}
\label{eq9da}
  A_2 &=& F+G \nonumber \\
  B_2 &=& i \frac{\Omega_2}{|\Omega_2|} (F-G)
\end{eqnarray}

In what follows we will consider $F$ and $G$ as first order quantities. To this order, from (\ref{ATeq3}), we have,
\begin{equation}\label{eq10d}
\gamma_+(t_+,r)= \gamma_0 + \kappa^2
\ln\left(\frac{r}{R_0}\right) - 2 \kappa \left(\psi_{in}(t_+,r)+\psi_{out}(t_+,r)\right)
\end{equation}
where $\kappa$ and $\gamma_0$ are given by (\ref{eq3a}). The important point here is that we have now expressions for $\psi_+$ and $\gamma_+$ where we have separate amplitudes for the incoming and the outgoing wave parts.

Notice that since (\ref{eq9d}) is just a reordering of terms in (\ref{eq4da}), all the previous relations regarding $t_{\pm}$, $\Omega$, and $\Omega_{\pm}$ are valid. Then replacing again the expressions for the dynamic variables, but using now (\ref{eq9d}) and (\ref{eq10d}), in (\ref{ATeq5}), (\ref{ATeq6}), and (\ref{ATeq9}), and
expanding to first order, we find this time a set of three linear independent
equations for $A_1$, $F$, $G$, and $\xi_0$.  In this paper we will be mainly interested in the evolution of an initially static shell, under the influence of a bounded incoming compact pulse. We choose therefore to solve these equations for  $A_1$, $G$, and $\xi_0$ in terms of $F$. The resulting explicit expressions are, unfortunately, rather long and difficult to read. Nevertheless, they have in general the structure,
\begin{eqnarray}
\label{13d1}
  A_1 &=& \frac{H_1} {D} F \nonumber \\
  G &=& \frac{H_2} {D} F \\
  \xi_0 &=& \frac{H_3} {D} F \nonumber
\end{eqnarray}
where $H_i\;,\; i =1,2,3$ are bounded regular functions of $(R_0,J,\Omega)$ and $D$ is given by,

\begin{eqnarray}
\label{13d2}
D & = &- \left[ 2\, \left( {R_{{0}}}^{2}+{J}^{2} \right) {J}^{2}R_{{0}}
 \left(  \left( 4\,{J}^{4}+6\,{R_{{0}}}^{2}{J}^{2}+{R_{{0}}}^{4}
 \right) {\Omega}^{2}-2\,{J}^{2} \right) J_{{0}} \left(  \left| \Omega
 \right| R_{{0}} \right) \right. \nonumber \\
& & \left. -\left( {R_{{0}}}^{2} \left( 2\,{J}^{2}+{R_{{0}}}^{2
} \right) ^{2} \left( {R_{{0}}}^{2}+{J}^{2} \right) ^{2}{\Omega}^{2}-{
J}^{2} \left( 2\,{R_{{0}}}^{6}+{J}^{2} \left( 2\,{J}^{2}+{R_{{0}}}^{2}
 \right) ^{2} \right)  \right) \Omega J_{{1}} \left(  \left| \Omega \right| R_{{0}
} \right)  \right] \nonumber \\
& &   \left(  \left| \Omega
 \right| J_{{0}} \left(  \left| \Omega_{{2}} \right| R_{{0}} \right) -
i\Omega\,Y_{{0}} \left(  \left| \Omega_{{2}} \right| R_{{0}} \right)
 \right) \nonumber \\
& & - \left[  \left( {R_{{0}}}^{2} \left( {R_{{0}}}^{2}+{J}^{2}
 \right) ^{2}{\Omega}^{2}-{J}^{2} \left( 2{R_{{0}}}^{2}+{J}^{2}
 \right)  \right) J_{{0}} \left(  \left| \Omega \right| R_{{0}}
 \right) \right. \nonumber \\
 & & \left.+2{J}^{2}  \Omega R_{{0}} \left( {R_{{0}}}^{2}+{J}^{2} \right)
 J_{{1}} \left(  \left| \Omega \right| R_{{0}}
 \right)
 \right] \Omega\, \left( 2\,{J}^{2}+{R_{
{0}}}^{2} \right) ^{2} \nonumber \\
& & \left( \Omega\,J_{{1}} \left(  \left| \Omega_{{2}} \right| R
_{{0}} \right) -i \left| \Omega \right| Y_{{1}} \left(  \left| \Omega_
{{2}} \right| R_{{0}} \right)  \right)
\end{eqnarray}
where $\Omega_2 = \Omega_+$.

In spite of its appearance it is not difficult to show that $D$ has no zeros for real $\Omega$. A zero of $D$ for real $\Omega$ would imply that the imaginary part of
\begin{equation}\label{14d}
\frac{\left| \Omega
 \right| J_{{0}} \left(  \left| \Omega_{{2}} \right| R_{{0}} \right) -
i\Omega\,Y_{{0}} \left(  \left| \Omega_{{2}} \right| R_{{0}} \right)}{ \Omega\,J_{{1}} \left(  \left| \Omega_{{2}} \right| R
_{{0}} \right) -i \left| \Omega \right| Y_{{1}} \left(  \left| \Omega_
{{2}} \right| R_{{0}} \right)}
\end{equation}
vanishes. But this is given by,
\begin{equation}\label{15d}
-\frac{J_0(|\Omega_2|R_0) Y_1(|\Omega_2|R_0)-J_1(|\Omega_2|R_0) Y_0(|\Omega_2|R_0)}{J_1(|\Omega_2|R_0)^2+ Y_1(|\Omega_2|R_0)^2} =
\frac{2}{\pi R_0 |\Omega_2| (J_1(|\Omega_2|R_0)^2+ Y_1(|\Omega_2|R_0)^2)}
\end{equation}
where we have used properties of the Bessel functions to obtain the right hand side, and, therefore, the imaginary part is non vanishing for all $\Omega$.

This result is crucial, because it means that if $F$ depends regularly on $\Omega$, the solutions obtained using (\ref{13d1}) and (\ref{13d2}) are well defined modes for all $\Omega$ in $-\infty < \Omega < +\infty$. Therefore, because of linearity, arbitrary linear combinations of these modes will also be solutions of the problem. Before discussing the possible completeness of this set of modes we will write explicit forms for the expressions we have in mind. These are given by,
\begin{equation}\label{16d}
 \psi_{in}(t_+,r)= \int_{-\infty}^{+\infty}F(\Omega) e^{i \Omega_2 t_+}\left[J_0(|\Omega_2|r)+i \frac{\Omega_2}{|\Omega_2|}Y_0(|\Omega_2|r)\right]  d \Omega
\end{equation}
where $F(\Omega)$ is an arbitrary complex function of $\Omega$ which we will assume at least square integrable.Then, taking into account (\ref{13d1}), we have for the remaining dynamical variables,
\begin{eqnarray}
\label{17d}
\psi_-(t_-,r)  &=& \int_{-\infty}^{+\infty} \frac{H_1}{D} F(\Omega) e^{i \Omega t_-} J_0(|\Omega|r) d \Omega  \nonumber \\
 \psi_{out}(t_+,r)  &=& \int_{-\infty}^{+\infty} \frac{H_2}{D} F(\Omega) e^{i \Omega_2 t_+}\left[J_0(|\Omega_2|r)-i \frac{\Omega_2}{|\Omega_2|}Y_0(|\Omega_2|r)\right]  d \Omega \nonumber  \\
  \xi(\tau) &=& \int_{-\infty}^{+\infty} \frac{H_3}{D} F(\Omega) e^{i \Omega \tau}  d \Omega
\end{eqnarray}

We shall analyze the properties of this expansion in the next Section, and show that they solve a well defined characteristic value problem for our system, and, therefore, in this sense, the mode expansion is also complete. We remark, for completeness, that (\ref{16d}), for real $\Omega$, represents the most general solution for $\psi_{in}(t_+,r)$, such that both $\psi_{out}(t_+,r)$ are bounded for large $r$.

\section{The characteristic value problem}

Consider again (\ref{16d}). The problem that we have in mind is one where at some given time the shell and a large neighbourhood of the space time that includes the shell and the symmetry axis, are in a static state, but there is a finite gravitational pulse incoming from large $r$. In other words, if $R_0$ corresponds to a static solution, then for some large time in the past and some $R_s >> R_0$, the solution in $0 \leq r \leq R_s$ coincides with the static solution. For $r > R_s$, on the other hand, there is region where $\psi_{in}(t_+,r)$ is non vanishing. To make this more precise, we recall that for large $r$ an incoming wave solution for $\psi_+(t_+,r)$ has the asymptotic form,
\begin{equation}\label{eq01e}
\psi_{in}(t_+,r) \simeq \frac{1}{\sqrt{r}} f(t_++r)
\end{equation}
where $f(u)$ is an arbitrary function of (essentially) compact support. We may compare this with the asymptotic form for large $r$ of (\ref{16d}),
\begin{equation}\label{eq02e}
\frac{1}{\sqrt{r}} f(t_++r) \simeq \int_{-\infty}^{+\infty}F(\Omega) \sqrt{\frac{1}{\pi r |\Omega_2|}}\left(1- i \frac{\Omega_2}{|\Omega_2|}\right)  e^{i \Omega_2 (t_++r)} d\Omega
\end{equation}
which can be inverted to give,
\begin{equation}\label{eq03e}
F \left( \Omega \right) = \frac{ |\Omega_2|^{3/2}}{2 \sqrt{\pi}(|\Omega_2|-i\Omega_2)}
\int_{-\infty}^{+\infty} f \left( u \right)   {\rm e}^{-i\Omega_2\,u}  du
\end{equation}

Therefore, we find a one--to--one relation between $F(\Omega)$ and characteristic data given for $r \to \infty$, $t_+ \to -\infty$. In this sense, we have proved the following result: {\em The set of modes given by (\ref{13d1}) and (\ref{13d2}), used in the construction of (\ref{16d}) and (\ref{17d}) is complete, and, in particular, (\ref{16d}) and (\ref{17d}) provide the evolution of the corresponding dynamical variables for arbitrary characteristic data given at $r \to \infty$, $t_+ \to -\infty$.}

In the next two Sections we shall use these results to analyze the stability of the static solutions and then consider the presence of quasi normal modes (QNM) and their associated quasi normal ringing (QNR).

\section{Stability of the static solutions}

Conceptually, we may say that a static configuration of the shell is stable if, given data arbitrarily close to that configuration, the system evolves towards that static configuration. Otherwise we would say that the static configuration is unstable. But, as we have already indicated, we have explicit forms for the evolution of the system when it is perturbed from an initially static configuration by an incoming gravitational pulse of (essentially) compact support. Each one of these pulses is uniquely characterized by a corresponding function $F(\Omega)$. Concentrating in particular  on $\xi(\tau)$, we have,
\begin{equation}\label{eq01ee}
H_3 = \frac{4\,i\Omega^2\,{R_{{0}}}^{7}}{\pi } \left({\frac {{J}^{2}   \left( 2\,{J}^{2}+3\,{R_{{0}}}^{2} \right) }{\Omega R_0 (2\,{J}^{2}+{R_{
{0}}}^{2})}}J_{{0}} \left(  \left| \Omega \right| R_{{0}}
 \right)  -\left( {R_{{0}}}^{2}
+{J}^{2} \right) J_{{1}} \left(  \left| \Omega \right| R_{{0}}
 \right)  \right)
\end{equation}
and, using (\ref{13d2}), we can check that for large $\Omega$,
\begin{equation}\label{eq02ee}
   \left| \frac{H_3}{D} \right| \sim \left|{\frac {2 \sqrt {2} {R_{{0}}}^{7/2} \cos \left( \Omega R_{{0}}+
\pi/4  \right) }{\sqrt {\pi } \left( 2 {J}^{2}+{R_{{0}}}^{2} \right)  \left( {R_{{0}}}^{2}
+{J}^{2} \right)   }}\right| \frac{1}{|\Omega|^{3/2} }
\end{equation}
plus terms of order $|\Omega|^{-5/2}$.
As consequence, on account of (\ref{17d}), and our assumptions on $F(\Omega)$, $\xi(\tau)$ is the Fourier transform of a square integrable function for {\em any} $J$ and $R_0$, but then, we must have $\xi(\tau) \to 0$ for $\tau \to +\infty$, and, therefore, {\em all } static configurations are stable. More explicitly, we have shown that all finite admissible gravitational pulses incoming from large $r$ can be described by an appropriate function $F(\Omega)$, and that this function is in one-to-one correspondence with the shape of the pulse at large $r$. We have also shown there is a unique evolution for the system corresponding to each function $F(\Omega)$. This evolution is such that the dynamic variables are essentially given by the Fourier transform of a square integrable functions and, therefore, they are themselves square integrable and must vanish in the limit $|\tau| \to \infty$. But this shows that all initially static configurations of the shell that are perturbed by
an incoming pulse of the type described above, will eventually settle back to their original
static configuration, and, in this sense, they are all {\em stable.}

Just for completeness, it should be clear that we have not found a self adjoint
extension associated to the coupled partial plus ordinary system of differential equations
of this problem, so that we could not state completeness of the mode expansion in the
same way as when such an extension is possible. What we have found is that the characteristic value
problem can indeed be solved by our mode expansion, since it reduces essentially to a Fourier transform, and, that �in
this sense�, the expansion is complete, and proves that the static configurations of the
system are stable under a perturbation that has the form of a bounded, (essentially) compactly
supported incoming pulse. But this is essentially all that is required to consider the system
�physically� stable. It is still an open and interesting question to find the relation, if it exists,
between the system of equations governing the evolution of the system and an associated
self adjoint problem that would be more useful an the analysis of the initial value problem.
Nevertheless, as show in Section 10, for the general initial plus boundary value problem we have a conserved, positive definite constant of the motion, and therefore we have stable evolution for any finite initial data.\\

We have already seen that for a shell with given values of $\lambda$ and $J$ there may be just one, two or no static solutions of the equations of motion. It has been suggested \cite{hamity} that in the case of two solutions only one is stable and the other is unstable. However, our present analysis does not show a qualitative difference in the behaviour of two static solutions under perturbations. There appears to be, nevertheless, a more subtle difference between the solutions when we analyze the evolution in more detail. This is related to the presence of quasi-normal modes, and is considered in the next Section.

\section{Quasi normal modes}

In this Section we analyze the presence of quasi normal modes and the associated quasi normal ringing in our system. Very roughly, QNM and QNR appear in systems coupled to a field that admits a decomposition in periodic incoming and outgoing waves.(See e.g. \cite{nollert} for more details, and further references). Generally QNM are associated to non trivial solutions where the incoming wave amplitude vanishes. This requires, generally, a complex value for the period, and therefore, in general, the corresponding outgoing and other amplitudes are unbounded in space and time, and, therefore, are not physical. In our problem, we have seen that for the periodic solutions, the dynamical variables other than $F(\Omega)$ are given by expressions of the form $H_i F/D$, and therefore, QNM, that is, nontrivial solutions with $F=0$, can exist only for values of $\Omega$ such that $D=0$. We have already seen that $D$ is non vanishing for {\em real} $\Omega$, but it may vanish for {\em complex} values of $\Omega$. These zeros of $D$ introduce complex poles in the expressions for $\xi$ and the other field variables whose complex frequency does not depend on $F$ although the amplitude of their possible contributions does depend on $F$. Since $\xi$, $A_1$ and $G$ are given by Fourier transforms of these expressions containing complex poles, we may get contributions dominated by these poles. These are the QNR amplitudes associated to the QNM. Therefore, to find the QNM we need to find the complex zeros of $D$. In principle this is the simple matter of solving the equation $D=0$ for $\Omega$. Unfortunately, if we consider the explicit expression for $D$ given by (\ref{13d2}), we notice that it contains Bessel functions of different types and arguments depending on $\Omega$, so that only a numerical computation of the zeros appears as feasible, but even in this case the accurate computation of these zeros is a complicated task. In this paper we only attempt a rather rough computation of the zeros closest to the real $\Omega$ axis in the complex $\Omega$ plane, as these would correspond to the most noticeable QNR of the system.

For this explicit computation we go back to (\ref{13d1}) and notice that if define $x=J/R_0$ and $\Sigma = \Omega R_0$, the equation $D=0$ may be written in the form,
\begin{eqnarray}
\label{eq01f}
  0 &=&  \left[ 2\, \left( 1+{x}^{2} \right) {x}^{2} \left( {\Sigma}^{2} \left( 4\,{x}^{4}+6\,{x}^{2}+1 \right) -2\,{x}^{2} \right) J_{{0}}
 \left(  \left| \Sigma \right|  \right) \right. \nonumber \\
 & & \left. -  \Sigma\, \left( {\Sigma}^{2} \left( 2\,{x}^{2}+1
 \right) ^{2} \left( 1+{x}^{2} \right) ^{2}-{x}^{2} \left( 4\,{x}^{4}+
2+{x}^{2}+4\,{x}^{6} \right)  \right) J_{{1}} \left(  \left| \Sigma
 \right|  \right) \right] \nonumber \\
& & \left( - \left| \Sigma
 \right| J_{{0}} \left(  \left| \Sigma \right|  \left( 2\,{x}^{2}+1
 \right) ^{2} \right) +i\Sigma\,Y_{{0}} \left(  \left| \Sigma \right|
 \left( 2\,{x}^{2}+1 \right) ^{2} \right)  \right)   \\
 & & - \left( \Sigma\,J_
{{1}} \left(  \left| \Sigma \right|  \left( 2\,{x}^{2}+1 \right) ^{2}
 \right) -i \left| \Sigma \right| Y_{{1}} \left(  \left| \Sigma
 \right|  \left( 2\,{x}^{2}+1 \right) ^{2} \right)  \right) \nonumber \\
 & & \left[   \left( {\Sigma}^{2}
 \left( 1+{x}^{2} \right) ^{2}-{x}^{2} \left( {x}^{2}+2 \right)
 \right)J_
{{0}} \left(  \left| \Sigma \right|  \right)  +2  \Sigma{x
}^{2} \left( 1+{x}^{2} \right) J_{{1}} \left(  \left| \Sigma \right|  \right) \right] \Sigma \left( 2{x}^{2}+1
 \right) ^{2} \nonumber
\end{eqnarray}

Actually, since we are looking for zeroes near the real $\Omega$ axis, we should make the replacements $|\Sigma|= \Sigma$ to look for zeros in the region $\Re(\Omega) > 0$, and $|\Sigma|= -\Sigma$ in the region $\Re(\Omega) < 0$.
In any case, the zeroes of $D$ satisfy the scale law,
\begin{equation}\label{eq02f}
    \Omega_{QNM}= \frac{1}{R_0} Q\left(\frac{J}{R_0}\right)
\end{equation}
where $Q$ is some complex function of its real argument.

We searched for zeroes of $D$ using two procedures. In the first we noticed that if we look for zeroes in the $\Re(\Sigma) > 0$ region we may replace
$|\Sigma|= \Sigma$ in (\ref{eq01f}), and after canceling some common factor of $\Sigma$, we find a maximum power  $\Sigma^3$ in the coefficients of the Bessel functions. We formally solve for this factor and obtain,
\begin{eqnarray}
\label{eq03f}
  \Sigma^3 &=&  \left[  \left( 2\, \left( 1+{x}^{2} \right)  \left(  \left( 4
\,{x}^{4}+6\,{x}^{2}+1 \right) {\Sigma}^{2}-2\,{x}^{2} \right) J_{{0}}
 \left( \Sigma \right) -J_{{1}} \left( \Sigma \right) \Sigma\, \left(
4\,{x}^{4}+2+{x}^{2}+4\,{x}^{6} \right)  \right)  \right. \nonumber \\
 & &  \left( J_{{0}}
 \left( \Sigma\, \left( 2\,{x}^{2}+1 \right) ^{2} \right) -iY_{{0}}
 \left( \Sigma\, \left( 2\,{x}^{2}+1 \right) ^{2} \right)  \right)  \nonumber \\
& &  -
 \left( 2\,J_{{1}} \left( \Sigma \right) \Sigma\, \left( 1+{x}^{2}
 \right) -J_{{0}} \left( \Sigma \right)  \left( {x}^{2}+2 \right)
 \right)  \nonumber \\
 & & \left. \left( iY_{{1}} \left( \Sigma\, \left( 2\,{x}^{2}+1 \right)
^{2} \right) -J_{{1}} \left( \Sigma\, \left( 2\,{x}^{2}+1 \right) ^{2}
 \right)  \right) \Sigma\, \left( 2\,{x}^{2}+1 \right) ^{2} \right] {x
}^{2}   \\
& & \left[ \left(  \left( iY_{{1}} \left( \Sigma\, \left( 2\,{x}^{2}+1
 \right) ^{2} \right) -J_{{1}} \left( \Sigma\, \left( 2\,{x}^{2}+1
 \right) ^{2} \right)  \right) J_{{0}} \left( \Sigma \right) \right. \right.  \nonumber \\
 && \left. \left.+ \left(
J_{{0}} \left( \Sigma\, \left( 2\,{x}^{2}+1 \right) ^{2} \right) -iY_{
{0}} \left( \Sigma\, \left( 2\,{x}^{2}+1 \right) ^{2} \right)
 \right) J_{{1}} \left( \Sigma \right)  \right)  \left( 2\,{x}^{2}+1
 \right) ^{2} \left( 1+{x}^{2} \right) ^{2} \right]^{-1} .\nonumber
\end{eqnarray}

We can use this equation in an iterative scheme where we input a value for $\Sigma$ on the right hand side, and the cubic root of the (complex) number obtained is inserted again as an ``improved'' value for $\Sigma$. We have numerically checked that this works very well for sufficiently small values of $x$, say $x < \sim 0.5$, but stops converging for larger values of $x$. A different, in a way more direct, although not very accurate method that we have used is to consider again (\ref{eq01f}), fix a value of $x$, and plot the real and the imaginary parts of (\ref{eq01f}), as functions of $\Re(\Sigma)$, for fixed $\Im(\Sigma)$. With this method one can easily visualize the possible common zeros of the curves and adjust, by trial an error, the best values of $\Re(\Sigma)$ and $\Im(\Sigma)$. We have found by this method that there are complex zeros of $D$ at least in the range $0 < x \sim 2$, and possibly for larger values of $x$. We have found zeros both near $\Re(\Sigma) > 0$ and $\Re(\Sigma) < 0$. Some of these are shown in Table I.\\

\begin{table}[h!]
\begin{tabular}{|c|c|c|}
   \hline
   x & $\Re(\Sigma) > 0$ & $\Re(\Sigma) < 0$\\
   \hline
   0.10 & 0.1396+ 0.00004i &- 0.1408+ 0.00015i  \\
   0.20 & 0.2685+ 0.00101i &- 0.2760+ 0.0035i \\
   0.30&  0.3772+ 0.00575i &- 0.3939+ 0.01775i \\
 0.40&  0.4612+ 0.01722i &- 0.4840+ 0.0479i \\
 0.50 & 0.5202+ 0.0366i & - 0.5410+ 0.0910i  \\
 0.60& 0.5570+ 0.0628i &- 0.5676+ 0.13998i \\
  0.70 & 0.5750+ 0.0938i & - 0.5704+ 0.1887i \\
  0.80 & 0.5780+ 0.127i &- 0.5552+ 0.2337i \\
   0.90 & 0.5704+ 0.16125i & - 0.5305+ 0.273i \\
   1.00& 0.5535+ 0.1943i &- 0.4970+ 0.307i \\
   1.20 & 0.5010+ 0.2542i & - 0.4200+ 0.3588i \\
    1.40 & 0.4350+ 0.3033i &- 0.3330+ 0.395i \\
    1.60 & 0.3590+ 0.342i & -0.2350+ 0.4201i \\
    1.80, & 0.2736+ 0.37175i & - 0.0960+ 0.4378i \\
    2.00 & 0.1708+ 0.3947i &  - - \\
   \hline
\end{tabular}
\caption{ The zeros of $D$.}
\end{table}

The first thing to notice is that all these zeroes have positive imaginary part, as one would expect from the stability arguments of the previous Section. We also notice that there are zeros both for $\Re(\Sigma) > 0$ and $\Re(\Sigma) < 0$, although having different imaginary parts and different $|\Re(\Sigma)|$. In general the imaginary part of $\Sigma$ increases with $x$, and the modes become very strongly damped for $x$ of the order or larger than one. All these results are in good agreement with the ``first class solutions'' of \cite{nakao} for $x<2$, although they obtain a much larger set of solutions. Regarding this point we remark that, unfortunately, the analytic structure in the complex $\Omega$ plane of the integrands in (\ref{17d}) is very complicated, and the exact relation between these zeroes and the evolution of the dynamical variables is not easily established. To obtain more information on the subject of QNR for our system we carried out numerical integrations of $\xi(\tau)$, as given by (\ref{17d}), for simple incoming $F$, and found examples of QNR, with parameters close to those of the zeros of $D$ found here. This is detailed in the next Section. Next, in Section IX, we consider again the perturbation problem in general, and obtain a set of coupled equations directly for $\xi$, $\psi_-$ and $\psi_+$, without resorting to a mode expansion. This allows us to set up an initial value problem that can be solved fully numerically, without the use of the mode expansion or Bessel functions.

\section{Numerical evaluation of the integrals in (\ref{17d})}

We have carried out explicit numerical integrations of (\ref{17d}) to obtain $\xi(\tau)$, with the choice,
\begin{equation}\label{eq01g}
F \left( \Omega_{{2}} \right) ={\frac {  \left| \Omega_{{2}}
 \right|^{3/2}} {\left(  \left| \Omega_{{2}}
 \right| -i\Omega_{{2}} \right)  }} \exp(-\Omega^2/6)
\end{equation}
This corresponds to an asymptotically incoming pulse of the form,
\begin{equation}\label{eq02g}
f(t_++r) = \sqrt{6}\exp(- \frac{3}{2}\left| t_++r \right|^2 )
\end{equation}

As a first choice we took $R_0=0.25$ and $J=0.125$. This corresponds to $x=0.5$ and, therefore, is within the range considered "stable" in previous analysis. The resulting $\xi(\tau)$ is displayed in Fig. 1. We can see the characteristic shape of a QNR, that is, a damped oscillation. As usual, we can get a better look at this shape by displaying, as in Fig. 2, $\alpha=\ln|\xi(\tau)|$ as a function of $\tau$ (thick line). We also display in Fig. 2 an approximate fit using the function $\ln(|\cos(\omega_r \tau +\varphi)|)-\omega_i \tau + q$, where $\omega_r=2.08$, $\varphi =0.01 $, $\omega_i=0.16$ and $q=4.28$, although in the actual figure, for clarity, we set $q=2$ to avoid superposition of the two graphs. These parameters are in good agreement with those corresponding to the QNM with $\Re(\Omega) > 0$ for $x=0.5$ shown in Table I.

\begin{figure}
\centerline{\includegraphics[height=12cm,angle=-90]{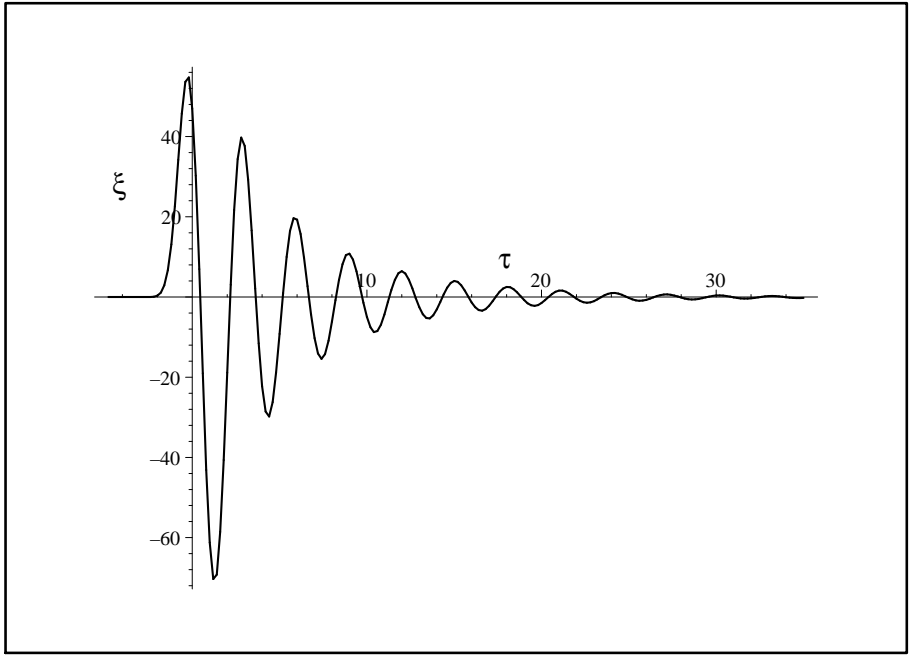}}
%\vspace{-1cm}
\caption{$\xi(\tau)$ as a function of $\tau$ for an incoming pulse of the form (\ref{eq02g}), for $R_0=0.25$ and $J=0.125$}
\end{figure}

\begin{figure}
\centerline{\includegraphics[height=12cm,angle=-90]{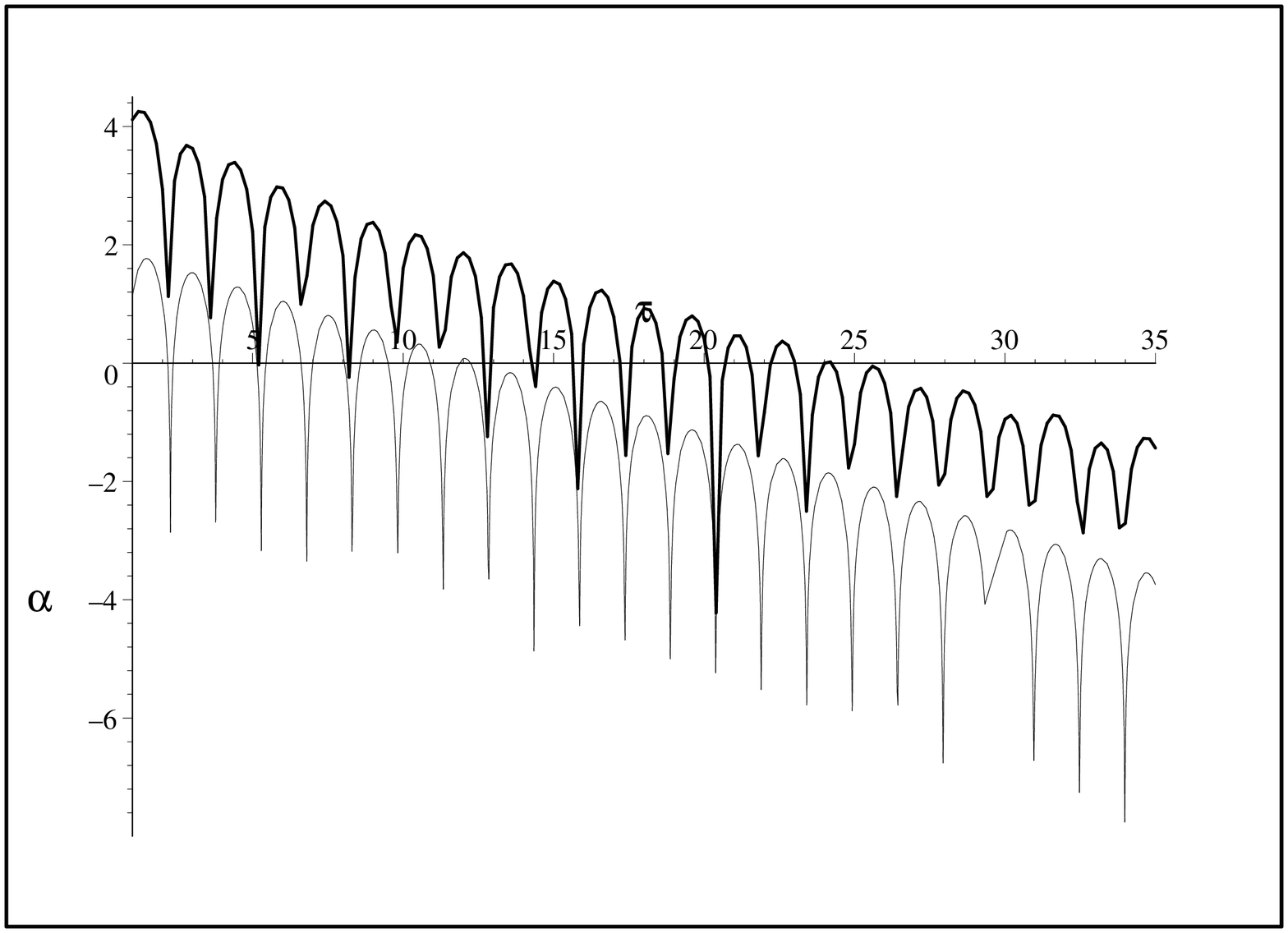}}
%\vspace{-1cm}
\caption{$\alpha = \ln|\xi(\tau)|$ for an incoming pulse of the form (\ref{eq02g}), for $R_0=0.25$ and $J=0.125$ (thick line). The thin line is an approximate fit by $\ln(|\cos(2.08 \tau +0.01)|)-0.16 \tau + 4.28$. Notice that the actual plot of this fit has been displaced down by 2.5 to avoid cluttering the figure. }
\end{figure}

As a second example we took $R_0=0.25$ and $J=0.25$. This corresponds to $x=1.0$ and, therefore, is outside the range considered "stable" in previous analysis. The resulting $\xi(\tau)$ is displayed in Fig. 3. We can see again the characteristic shape of a QNR, but, in this case, as a strongly damped oscillation. This result is also in good agreement with the fact that the QNM for $x=1.0$ shown in Table I corresponds to a strongly damped oscillation. We remark once again that the analytic structure of the integrand in (\ref{17d}) is far from simple, and that although several sets of complex QNM frequencies can be established rather accurately, their effect on the resulting amplitudes cannot be easily established. Nevertheless, the integrals in (\ref{17d}) are taken along the real $\Omega$ axis, and therefore, they are insensitive to the manner in which the integrand might be extended to, e.g., the upper complex-$\Omega$ plane, where the poles corresponding to the QNM reside.

\begin{figure}
\centerline{\includegraphics[height=12cm,angle=-90]{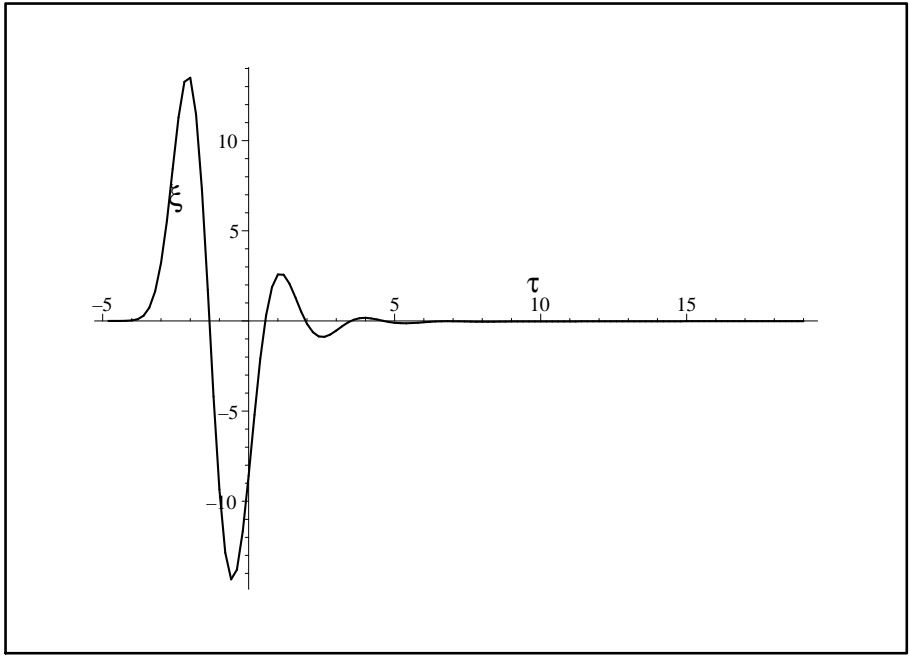}}
%\vspace{-1cm}
\caption{$\xi(\tau)$ as a function of $\tau$ for an incoming pulse of the form (\ref{eq02g}), for $R_0=0.25$ and $J=0.25$, that is $x=1.0$. Notice the strong damping of the oscillations after the arrival of the incoming pulse.}
\end{figure}

In the next Section we consider a different approach, where we linearize the equations of motion and obtain a set of coupled linear equations for the amplitudes, together with a set of boundary and matching conditions that make possible to establish (and numerically solve) a well defined initial data problem.

\section{The general perturbation problem}

We consider again the set of dynamical equations and matching conditions for our problem, assuming that we are close to a static solution, characterized by certain values of $J$ and $R_0$. We therefore write,
\begin{eqnarray}
\label{eq01h}
  R(\tau) &=& R_0 + \epsilon \; \xi(\tau) \nonumber \\
  \psi_-(t_-,r) &=& \epsilon\; \chi_1(t_-,r) \\
  \psi_+(t_+,r) &=& - \kappa \ln (r/R_0) + \epsilon \;\chi_2(t_+,r) \nonumber
\end{eqnarray}
with $\chi_i$ satisfying the equations,
\begin{eqnarray}
\label{eq01ha}
 \frac{\partial^2 \chi_1}{\partial t_-^2} -\frac{\partial^2 \chi_1}{\partial r^2}-\frac{1}{r}\frac{\partial \chi_1}{\partial r} &=& 0  \\
 \label{eq01hb}
 \frac{\partial^2 \chi_2}{\partial t_+^2} -\frac{\partial^2 \chi_2}{\partial r^2}-\frac{1}{r}\frac{\partial \chi_2}{\partial r} &=& 0
\end{eqnarray}
and expand the remaining equations to first order in $\epsilon$. Using these expansion we first obtain,
\begin{eqnarray}
\label{eq02h}
  \gamma_-(t_-,r) &=& {\cal{O}}(\epsilon^2)   \\
  \gamma_+(t_+,r) &=& \gamma_0 +\kappa^2 \ln (r/R_0) - 2 \epsilon\;\kappa \;\chi_2(t_+,r) \nonumber
\end{eqnarray}
In these equations $\kappa$ and $\gamma_0$ are given by (\ref{eq3a}) and (\ref{eq4a}), so that the zeroth order terms coincide with the static solution of Section III. From these, again to first order we find,
\begin{eqnarray}
\label{eq03h}
 \frac{ dt_-}{d\tau} &=& 1 + {\cal{O}}(\epsilon)    \\
 \frac{ dt_+}{d\tau} &=& e^{-\gamma_0} + {\cal{O}}(\epsilon) \nonumber
\end{eqnarray}
The explicit form of the terms of order $\epsilon$ is not required in what follows. Choosing appropriately some integration constants we set,
\begin{eqnarray}
\label{eq04h}
 t_-(\tau) &=& \tau + {\cal{O}}(\epsilon)    \\
 t_+(\tau) &=& e^{-\gamma_0} \tau + {\cal{O}}(\epsilon) \nonumber
\end{eqnarray}

Next we replace in the matching conditions and expanding again to first order we obtain two equations for $\chi_1$ and $\chi_2$ on the boundary $r=R_0$
\begin{eqnarray}
\label{eq05h}
0 & = & {\frac {2 {J}^{4} \left( 6\,{R_{{0}}}^{2}{J}^{2}+4{J}^{4}+3{R_{{0
}}}^{4} \right)   }{ \left( {R_{{0}}}^{2}+{J}^{2
} \right)  \left( 2{J}^{2}+{R_{{0}}}^{2} \right) ^{2}{R_{{0}}}^{4}}} \xi \left( \tau \right)  \nonumber \\
& &
-{\frac {2 R_{{0}}{J}^{4}  }
{ \left( {R_{{0}}}^{2}+{J}^{2} \right)  \left( 2{J}^{2}+{R_{{0}}}^{2
} \right) ^{2}}} \chi_1 \left( \tau,R_{{0}} \right)
- \left.{\frac {\partial \chi_1 \left(\tau,r \right)}{\partial r}}\right|_{r=R_0}
 \\
 & &
-{\frac {2{J}^{2} \left( 4{J}^{2}+{R_{{0}}}^{2}
 \right)   }{ \left( 2{J}^{
2}+{R_{{0}}}^{2} \right) ^{2}R_{{0}}}}\chi_2 \left( e^{-\gamma_0}\tau,R_{{0}} \right)
+{\frac {   {R_{{0}}}^{4}}{ \left( 2\,{J}^{2}+{R_{{0}}}^{2}
 \right) ^{2}}}
 \left.{\frac {\partial \chi_2 \left(e^{-\gamma_0}\tau,r \right)}{\partial r}}\right|_{r=R_0}, \nonumber \\
0 & = & {\frac {2 {J}^{2}  }{{R_{{0}}}^{3}}}\xi \left( \tau \right)
+\chi_1 \left( \tau,R_{{0}} \right)
-\chi_2 \left(e^{-\gamma_0}\tau,R_{{0}}
 \right) \nonumber
\end{eqnarray}
and an equation of motion for $\xi(\tau)$,
\begin{eqnarray}
\label{eq06h}
{\frac {d^{2}}{d{\tau}^{2}}}\xi \left( \tau
 \right) & = & -{\frac {{J}^{2} \left( 4{J}^{6}+6{R_{{0}}}^{2}{J}^{4}+5{J}^{2}{R
_{{0}}}^{4}+2{R_{{0}}}^{6} \right)  }{
 \left( 2\,{J}^{2}+{R_{{0}}}^{2} \right) ^{2} \left( {R_{{0}}}^{2}+{J}
^{2} \right) ^{2}{R_{{0}}}^{2}}}  \xi \left( \tau \right) \nonumber\\
& & +{\frac {{J}^{2}{R_{{0}}}^{3} \left( 3
{J}^{2}+2{R_{{0}}}^{2} \right)   }{ \left( 2{J}^{2}+{R_{{0}}}^{2} \right) ^{2} \left( {R_{{0
}}}^{2}+{J}^{2} \right) ^{2}}} \chi_1 \left( \tau,R_{{0}}
 \right)
 +{\frac {   {R_{{0}}}^{2}}{{R_
{{0}}}^{2}+{J}^{2}}}
\left.{\frac {\partial \chi_{{1}} \left( \tau,r \right)}{\partial r }}
 \right|_{r=R_0}  \\
 & &
+{\frac { \left( 4{J}^{2}+{R_{{0}}}^{2
} \right) R_{{0}}{J}^{2}  }{
 \left( {R_{{0}}}^{2}+{J}^{2} \right)  \left( 2{J}^{2}+{R_{{0}}}^{2}
 \right) ^{2}}}\chi_2 \left( e^{-\gamma_0}\tau,R_{{0}} \right) \nonumber
\end{eqnarray}

For explicit numerical integration it turns out to be convenient to introduce instead of $\chi_2(t_+,r)$ a new function $\chi_3(\tau,r)$ defined by,
\begin{equation}\label{eq07h}
\chi_3(\tau,r)=\chi_2(e^{-\gamma_0}\tau,r)
\end{equation}
and, therefore, satisfying the equation:
\begin{equation}\label{eq08h}
-\frac{(2 J^2+R_0^2)^4}{R_0^8} \frac{\partial^2 \chi_3(\tau,r)}{\partial \tau^2}+\frac{\partial^2 \chi_3(\tau,r)}{\partial r^2}+\frac{1}{r}\frac{\partial \chi_3(\tau,r)}{\partial r} = 0
\end{equation}
This way all the dynamic variables involved evolve directly in terms of $\tau$. This simplifies the numerical treatment of the problem as it eliminates the need of separate grid spacing for $r<R_0$ and $r>R_0$. But, most important, using $\chi_3$ we may demonstrate the existence of a crucial constant of the motion, as is shown in the next Section.

\section{A positive definite constant of the motion.}

As discussed in the previous Section, we have a complete solution for characteristic data, and we can use this fact to prove stability regarding the evolution resulting from that type of data. This, nevertheless, leaves still open the question of the evolution of general initial data. In other words, the possibility of the existence of initial data that somehow is not registered as characteristic data, but such that it renders the motion unstable, in the sense that, i.e. $\xi(\tau)$ can acquire arbitrarily large values starting from finite initial data. To answer this question we consider again (\ref{eq05h}), replace $\chi_2$ in terms of $\chi_3$, and solve for $\chi_1(\tau,R_0)$ and $\chi_3(\tau,R_0)$. We get,

\begin{eqnarray}
\label{stab01}
\chi_{{1}} \left( \tau,R_0 \right) & = & -{\frac {R_0\,
 \left( {R_0}^{2}+{J}^{2} \right)  \left( 2\,{J}^{2}+{R_0}^{
2} \right) ^{2}} {2 {J}^{2} \left( 6\,{R_0}^{2}{J}^{2}+4\,{J}^{4}+{{\it R0
}}^{4} \right) }}
\left.\left({\frac {\partial }{\partial r}}\chi_{{1}} \left( \tau,r
 \right) \right) \right|_{r=R_0} \nonumber \\
& &+{\frac {{R_0}^{5} \left( {R_0}^{2}+{J
}^{2} \right) } {2{J}^{2} \left( 6\,{R_0}^{2}{J}^{2}+4\,{J}^{4}+{{\it R0
}}^{4} \right) }}
\left.\left({\frac {\partial }{\partial r}}\chi_{{3}} \left( \tau,r
 \right) \right) \right|_{r=R_0} \\
& &-{\frac {{J}^{2}  \left( 4\,{{
\it R0}}^{2}{J}^{2}+4\,{J}^{4}-{R_0}^{4} \right) }{{R_0}^{3}
 \left( 6\,{R_0}^{2}{J}^{2}+4\,{J}^{4}+{R_0}^{4} \right) }} \xi \left( \tau \right) \nonumber
\end{eqnarray}
and,
\begin{eqnarray}
\label{stab02}
 \chi_{{3}} \left( \tau,R_{{0}} \right)& = & -{\frac {R_0\,
 \left( {R_0}^{2}+{J}^{2} \right)  \left( 2\,{J}^{2}+{R_0}^{
2} \right) ^{2}} {2 {J}^{2} \left( 6\,{R_0}^{2}{J}^{2}+4\,{J}^{4}+{{\it R0
}}^{4} \right) }}
\left.\left({\frac {\partial }{\partial r}}\chi_{{1}} \left( \tau,r
 \right) \right) \right|_{r=R_0} \nonumber \\
& &+{\frac {{R_0}^{5} \left( {R_0}^{2}+{J
}^{2} \right) } {2{J}^{2} \left( 6\,{R_0}^{2}{J}^{2}+4\,{J}^{4}+{{\it R0
}}^{4} \right) }}
\left.\left({\frac {\partial }{\partial r}}\chi_{{3}} \left( \tau,r
 \right) \right) \right|_{r=R_0} \\
& &+{\frac {{J}^{2}   \left( 3\,{R_{
{0}}}^{2}+2\,{J}^{2} \right)  \left( 2\,{J}^{2}+{R_{{0}}}^{2} \right)
}{{R_{{0}}}^{3} \left( 6\,{R_{{0}}}^{2}{J}^{2}+4\,{J}^{4}+{R_{{0}}}^{4
} \right) }} \xi \left( \tau \right)
 \nonumber
\end{eqnarray}

Replacing now (\ref{stab01}) and (\ref{stab02}) in (\ref{eq06h}), upon multiplication by $ d\xi/d\tau$, and some rearrangement, we find,

\begin{eqnarray}
\label{stab03}
0 & = & \frac{d \xi(\tau)}{d\tau}{\frac {d^{2}\xi \left( \tau \right) }{d{\tau}^{2}}}
+{\frac {2 {J}^{
2}  }{6\,{R_{{0}}}^{2}{J}^{2}+4\,{J}^{4}+{R_{{0}
}}^{4}}} \xi \left( \tau \right) \frac{d \xi(\tau)}{d\tau} \nonumber \\
& &-{\frac {{R_{{0}}}^{2} \left( 4\,{R_{{0}}}^{2}{J}^{2}+4\,
{J}^{4}-{R_{{0}}}^{4} \right)  }{ 2 \left( {R_{{0}}}^{2}+{J}^{2} \right)
 \left( 6\,{R_{{0}}}^{2}{J}^{2}+4\,{J}^{4}+{R_{{0}}}^{4} \right) }}\frac{d \xi(\tau)}{d\tau} \left.\left({\frac {\partial }{\partial r}}\chi_{{1}} \left( \tau,r
 \right) \right) \right|_{r=R_0}  \\
 & & -{\frac { \left( 3\,{R_{{0}}}^{2}+2\,{J}^{2} \right) {R_{{0}}}^{6}  }{
 2\left( {R_{{0}}}^{2}+{J}^{2} \right)  \left( 6\,{R_{{0}}}^{2}{J}^{2}+
4\,{J}^{4}+{R_{{0}}}^{4} \right)  \left( 2\,{J}^{2}+{R_{{0}}}^{2}
 \right) }} \frac{d \xi(\tau)}{d\tau}\left.\left({\frac {\partial }{\partial r}}\chi_{{3}} \left( \tau,r
 \right) \right) \right|_{r=R_0}
\nonumber
\end{eqnarray}

We notice that the first line (\ref{stab03}) has the form of a total $\tau$-derivative. We may put the whole expression in this form as follows. We first take the derivative of (\ref{stab01}) with respect to $\tau$, solve for $ d\xi/d\tau$ and replace it in the second line in (\ref{stab03}). Next, we derive (\ref{stab02}) with respect to $\tau$, solve again for $ d\xi/d\tau$, and replace it in the third line in (\ref{stab03}). After a new rearrangement we get,

\begin{eqnarray}
\label{stab04}
0 & = &  {\frac {d \xi}{d\tau}} {\frac {d^{2}\xi }{d{\tau}^{2}}}
+{\frac {2 {J}^{2}  }{6\,{R_{{0}}}^{2}{J}^{2}+4\,{J}^{4}+{R_{{0}}}^{4}}} \xi {\frac {d \xi}{d\tau}} \nonumber \\
& &+{\frac {{R_{{0}}}^{14} \left(  \dfrac{\left( 2\,{J}^{2}+{R_{{0}}}^{2} \right) ^{2} }{R_0^4} {\dfrac {\partial ^{2}\chi_{{1}}}{
\partial \tau\partial r}}
  -{\dfrac {
\partial ^{2}\chi_{{3}}}{\partial \tau\partial r}} \right)
\left( \dfrac{ \left( 2\,{J}^{2}+{R_{{0}}}^{2} \right) ^{2
}}{R_0^4}{\dfrac {\partial \chi_{{1}}}{\partial r}}  -
   {\dfrac {\partial \chi_{{3}}}{\partial r}}    \right) }{4 \left( 2\,{J}^{2}+{R_{{0}}}
^{2} \right) ^{2}{J}^{4} \left( 6\,{R_{{0}}}^{2}{J}^{2}+4\,{J}^{4}+{R_
{{0}}}^{4} \right) }}
 \\
& & +{\frac {{R_{{0}}}^{5}  }{ 2\left( {R_{{0}}}^{2}+{J}^{2} \right) {J}^{2}}}
 {\frac {\partial \chi_{{1}}}{\partial r}} {\frac {\partial \chi_{{1}}}{\partial \tau}}
-{\frac {{R_{{0}}}^{9}   }{ 2\left( {R_{{0}}}^{2}+{J}^{
2} \right)  \left( 2\,{J}^{2}+{R_{{0}}}^{2} \right) ^{2}{J}^{2}}}
 {\frac {\partial \chi_{{3}}}{\partial r}}  {\frac {\partial \chi_{{3}}}{\partial \tau
}}  \nonumber
\end{eqnarray}
where, again, all derivatives of $\chi_1$ and $\chi_3$ are evaluated at $r=R_0$.

The first two lines in (\ref{stab04}) are total $\tau$-derivatives. To analyze the last line we notice that, on account of (\ref{eq01ha}) we have,
\begin{eqnarray}\label{stab05}
    \frac{d}{d\tau} \int_0^{R_0}\frac{r}{2}\left[\left(\frac{\partial \chi_1}{\partial \tau}\right)^2 +\left(\frac{\partial \chi_1}{\partial r}\right)^2 \right] dr & = &\left.\left( r  \frac{\partial \chi_1(\tau,r)}{\partial \tau}\frac{\partial \chi_1(\tau,r)}{\partial r}\right) \right|_0^{R_0} \nonumber \\
    & = & \left. R_0 \frac{\partial \chi_1(\tau,R_0)}{\partial \tau}\;\frac{\partial \chi_1(\tau,r)}{\partial r} \right|_{r=R_0}
\end{eqnarray}
where in the last line we have assumed that $\chi_{1}$ satisfies the (regularity) boundary conditions that $\chi_1(\tau,0)$ is finite, and that $(\partial \chi_1/\partial r) |_{r=0}=0$.

Similarly, from (\ref{eq08h}), we have,
\begin{eqnarray}\label{stab06}
    \frac{d}{d\tau} \int_{R_0}^{\infty}\frac{r}{2}\left[\frac{(2 J^2+R_0^2)^4}{R_0^8}\left(\frac{\partial \chi_3}{\partial \tau}\right)^2 +\left(\frac{\partial \chi_3}{\partial r}\right)^2 \right] dr & = &\left. \left( r  \frac{\partial \chi_3(\tau,r)}{\partial \tau} \frac{\partial \chi_3(\tau,r)}{\partial r}\right) \right|_{R_0}^{\infty} \nonumber \\
    & = & -\left. R_0 \frac{\partial \chi_3(\tau,R_0)}{\partial \tau}\;\frac{\partial \chi_3(\tau,r)}{\partial r} \right|_{r=R_0}
\end{eqnarray}
where we have assumed that $\psi_3(\tau,r)$ is such that the integral exists, and, therefore, there is no contribution for $r \to \infty$.

Comparing (\ref{stab05}) and (\ref{stab06}) with (\ref{stab04}) we find that, if $\psi_1(\tau,r)$ corresponds to a solution of (\ref{eq08h}) that is regular for $r=0$, (and, therefore, $r (\partial \psi_1/\partial r) (\partial \psi_1/\partial \tau) \to 0$ for $r \to 0$), and $\psi_3(\tau,r)$ is such that
$r (\partial \psi_3/\partial r) (\partial \psi_3/\partial \tau) \to 0$ for $r \to \infty$, (which must happen for the integral in (\ref{stab06}) to exist), then the quantity,
\begin{eqnarray}
\label{stab07}
E_s & = &  \frac{1}{2}\left(\frac {d \xi}{d\tau}\right)^2
+{\frac {{J}^{2}  }{6\,{R_{{0}}}^{2}{J}^{2}+4\,{J}^{4}+{R_{{0}}}^{4}}} \xi^2\nonumber \\
& &+{\frac {{R_{{0}}}^{14}   }{8 \left( 2{J}^{2}+{R_{{0}}}
^{2} \right) ^{2}{J}^{4} \left( 6{R_{{0}}}^{2}{J}^{2}+4{J}^{4}+{R_
{{0}}}^{4} \right) }}
\left. \left( \dfrac{ \left( 2{J}^{2}+{R_{{0}}}^{2} \right) ^{2
}}{R_0^4}{\dfrac {\partial \chi_{{1}}}{\partial r}}  -
   {\dfrac {\partial \chi_{{3}}}{\partial r}}\right)^2 \right|_{r=R_0} \nonumber
 \\
& & +{\frac {{R_{{0}}}^{4}  }{ 2\left( {R_{{0}}}^{2}+{J}^{2} \right) {J}^{2}}}
 \int_0^{R_0}\frac{r}{2}\left[\left(\frac{\partial \chi_1}{\partial \tau}\right)^2 +\left(\frac{\partial \chi_1}{\partial r}\right)^2 \right] dr \\
& & +{\frac {{R_{{0}}}^{8}   }{ 2\left( {R_{{0}}}^{2}+{J}^{
2} \right)  \left( 2\,{J}^{2}+{R_{{0}}}^{2} \right) ^{2}{J}^{2}}}
 \int_{R_0}^{\infty}\frac{r}{2}\left[\frac{(2 J^2+R_0^2)^4}{R_0^8}\left(\frac{\partial \chi_3}{\partial \tau}\right)^2 +\left(\frac{\partial \chi_3}{\partial r}\right)^2 \right] dr  \nonumber
\end{eqnarray}
is a {\em constant of the motion}, i.e., $d E_s/d\tau=0$. Notice that $E_s$ is positive definite for any non trivial solution of the equations of motion. Therefore, it provides an absolute bound for each term in (\ref{stab07}). Thus, if at any time $\tau$ all the terms in (\ref{stab07}) are finite, then they will remain finite for any evolution that satisfies the condition of regularity for $r=0$, demonstrating the absolute stability of the system under any finite perturbation for which all the terms in (\ref{stab07}) exist. Of course, a simple example is the case of an incoming wave of (essentially) compact support, for which we have already demonstrated the stability.

We should further notice that the term,
\begin{equation}\label{stab08}
T_d=\left. \left( \dfrac{ \left( 2{J}^{2}+{R_{{0}}}^{2} \right) ^{2
}}{R_0^4}{\dfrac {\partial \chi_{{1}}}{\partial r}}  -
   {\dfrac {\partial \chi_{{3}}}{\partial r}}\right)^2 \right|_{r=R_0}
\end{equation}
in the second line of (\ref{stab07}) allows for mild compensating divergences of $\partial \chi_1/\partial r$, and $\partial \chi_3/\partial r$, provided they are compatible with the existence of the integrals in (\ref{stab07}). We notice nevertheless that, taking into account (\ref{stab01}), and (\ref{stab02}), we may also write $T_d$ in the alternative forms,
\begin{eqnarray}
\label{stab09}
  T_d &=& \frac{4{J}^{4} \left( 6{R_{{0}}}^{2}{J}^{2}+4{J}^{4}+{R_{{0}}}^{4}
 \right) ^{2}}  {{R_{{0}}}^{10}
 \left( {R_{{0}}}^{2}+{J}^{2} \right) ^{2}}
 \left( \chi_{{1}} \left( \tau,R_{{0}} \right) +{\frac {{
J}^{2}   \left( 4{R_{{0}}}^{2}{J}^{2}+4{J}^{
4}-{R_{{0}}}^{4} \right) }{{R_{{0}}}^{3} \left( 6\,{R_{{0}}}^{2}{J}^{2
}+4{J}^{4}+{R_{{0}}}^{4} \right) }}\xi \left( \tau \right) \right) ^{2}
\\
   &=& \frac{4{J}^{4} \left( 6\,{R_{{0}}}^{2}{J}^{2}+4{J}^{4}+{R_{{0}}}^{4}
 \right) ^{2}} {  {R_{{0}}}^{10} \left( {R_{{0}}}^{2}+{J}^{2} \right) ^{2}}
 \left( {\frac {{J}^{2}   \left( 8R_0^2J^2+3R_0^4+4J^4
 \right)
 }{{R_{{0}}}^{3} \left( 6{R_{{0}}}^{2}{J}^{2}+4{J}^{4}+{R_
{{0}}}^{4} \right) }}\xi \left( \tau \right)
-\chi_{{3}} \left( \tau,R_{{0}} \right)  \right)
^{2} \nonumber
\end{eqnarray}
and, since both $T_d$ and $\xi(\tau)$ are bounded, we find that both $\chi_1(\tau,R_0)$ and $\chi_3(\tau,R_0)$ must also remain bounded, and this implies that $\chi_2(t,r)$ is also bounded. On account of (\ref{eq02h}), this result implies that $\gamma_+(t,r)$ is also bounded to first order. We remark that $\gamma_-$ is directly of second order and vanishes to first order. Therefore, the full metric remains bounded under arbitrary perturbations.

\section{Numerical solutions of the initial plus boundary conditions problem}

One can now check that the system (\ref{eq01ha}), (\ref{eq05h}), (\ref{eq06h}), and (\ref{eq08h}), with appropriate replacements of $\chi_2$ by $\chi_3$, can be used to set up and solve numerically the initial plus boundary value problem where, for some arbitrarily chosen $\tau=\tau_0$ we give arbitrary values to $\xi(\tau_0)$, $(d\xi(\tau)/d \tau)|_{\tau=\tau_0}$, to $\chi_1(\tau_0,r)$ and $(\partial \chi_1(\tau,r)/ \partial \tau)|_{\tau=\tau_0}$, in the interval $0 \leq r \leq R_0$, and to $\chi_3(\tau_0,r)$ and $(\partial \chi_3(\tau,r)/ \partial \tau)|_{\tau=\tau_0}$, in the interval $R_0 \leq r $, subject to the constraints $\partial \chi_1(\tau,0)/\partial \tau = 0$, and those resulting from (\ref{eq05h}).

Although this procedure is well defined, and leads to a unique evolution for appropriate
initial data, at least in its Cauchy domain, we must remark that the well-posedness of this
type of problems, as regards sensitivity to small changes in the initial data, is still an open and interesting issue, outside the limits of the present
research. The purpose of this exercise is mainly to display the evolution of some appropriately chosen initial data, in particular of $\xi(\tau)$, in the region where we may expect the presence (or absence) of quasi normal ringing. As we shall see, this behaviour, at least in the region analyzed, is in complete agreement with our previous results.

The integration procedure we have chosen uses a simple finite difference method for updating $\chi_1$ in $0\leq r <R_0$, and for $\chi_3$ in $R_0 < r \leq r_o$, where $r_o >>R_0$ is some appropriately chosen outer boundary. The values of $\xi$ are updated using a simple leap frog scheme. Once these updates are carried out, we use the matching conditions (\ref{eq05h}) to update $\chi_1$, and $\chi_3$ at $r=R_0$.
%Old: The integration is carried out from $\tau=0$ to $\tau \sim r_o/2$, so that the choice of boundary condition at $r=r_o$ is irrelevant.
%New:
We will be mainly interested in the behaviour of $\xi$ as a function of $\tau$. On this account we choose some appropriate value of $r_o$ and boundary condition for $\chi_3$ at $r_o$, but carry out the integration only from $\tau=0$ to $\tau \sim r_o/2$, to ensure, by causality, that no signal coming in from $r=r_o$ has time to reach $r=R_0$ and affect that behaviour. In particular, in all the examples below we set $\chi_3(r_o,t)=0$, while we chose $r_o=50$ for the first and second example, and $r_o=300$ for the other two examples.

The first explicit example of the results of this numerical integration for $\xi(\tau)$ is given in Fig.4 and Fig. 5. The initial data used was:
\begin{eqnarray}
\label{eq09h}
  \chi_3(0,r) &=& e^{-50(r-1)^2}  \;\;;\;\;
  \left.\frac{\partial \chi_3} {\partial \tau}\right|_{\tau=0} = -\frac{100 R_0^4 (r-1)}{(2 J^2 + R_0^2)^2} e^{-50(r-1)^2} \nonumber \\
 \chi_1(0,r) &=& 0 \;\;;\;\;
  \left.\frac{\partial \chi_1}{\partial \tau}\right|_{\tau=0} = 0  \\
   \xi(0) &=& 0 \;\;;\;\;
  \left.\frac{ d \xi}{d \tau}\right|_{\tau=0} = 0 \nonumber
\end{eqnarray}
and we set $R_0=0.25$, $J=0.125$, corresponding to $x=0.5$, and $r_o=50$.
In Fig. 4 we display the evolution of $\xi(\tau)$ as a function of $\tau$. The graph corresponds clearly to an evolution dominated by QNR, although a more detailed analysis indicates the presence of a small non oscillating background at large $\tau$. In Fig 5 we have a plot of $\ln(|\xi(\tau)|)$ (thick line) in the region dominated by the damped oscillation mode. The thin line curve corresponds to an approximate fit with the function $\ln(|\cos(2.18 \tau -1.4)|)-0.147 \tau $, although the actual plot in Fig. 5 has been displaced up by 0.5 for clarity. We can see again the very clear signal of a damped oscillation, with a frequency and fall off in very good agreement with those of the example of Fig. 2, although we have used a different procedure and form for the incoming pulse. The agreement might probably improve using a more elaborate rather than our simple minded implementation of the numerical procedure. In any case, it was not the purpose of the authors to optimize it but rather to show that even with a simple implementation we can solve the initial value problem related to the perturbation treatment of the dynamics of our system close to a static configuration, and obtain results in perfect agreement of those extracted from the mode expansion.

\begin{figure}
\centerline{\includegraphics[height=12cm,angle=-90]{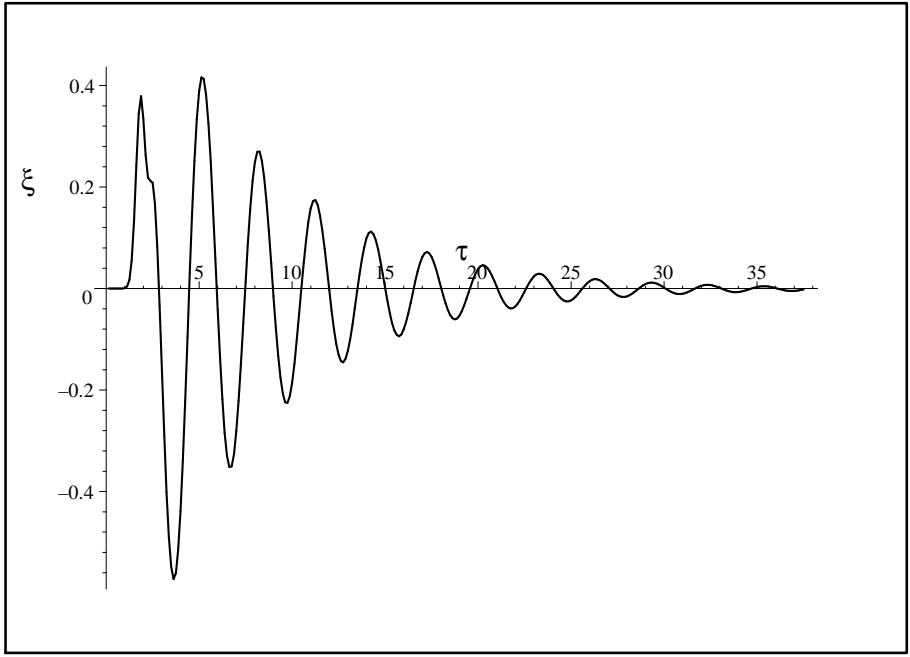}}
%\vspace{-1cm}
\caption{$\xi(\tau)$ as a function of $\tau$ for initial data of the form (\ref{eq09h}), for $R_0=0.25$ and $J=0.125$, that is $x=0.5$.}
\end{figure}

\begin{figure}
\centerline{\includegraphics[height=12cm,angle=-90]{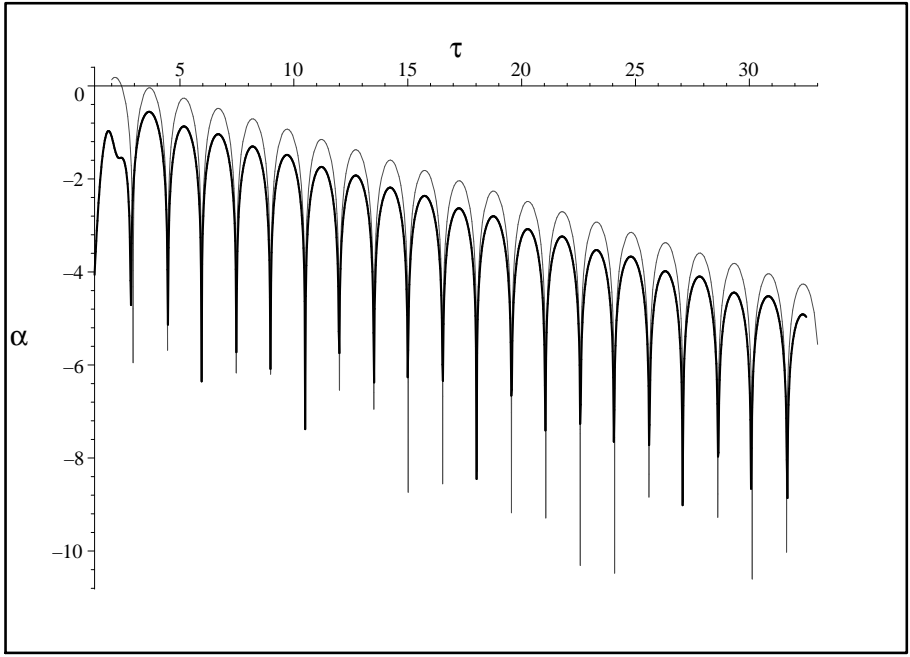}}
%\vspace{-1cm}
\caption{$\alpha = \ln|\xi(\tau)|$ for initial data of the form (\ref{eq09h}), for $R_0=0.25$ and $J=0.125$ (thick line). The thin line is an approximate fit by $\ln(|\cos(2.08 \tau -1.4)|)-0.147 \tau$. Notice that the actual plot of this fit has been displaced up by 0.5 to avoid cluttering the figure.}
\end{figure}

As a second full numerical example we considered again the initial data given (\ref{eq09h}), but setting $R_0=0.25$ and $J=0.25$, that is $x=1.0$. The results are shown in Fig. 6. In this case, in agreement with our analysis of the QNM, we find a very strongly damped oscillation, very much like that shown in Fig 3. for the same value of the parameters $R_0$ and $J$, although the computational procedures applied in each case were completely different. We can also check that both the frequency and damping are in good agreement with the values given in Table I for $x=1.0$ and positive real part for $\Omega$.

\begin{figure}
\centerline{\includegraphics[height=12cm,angle=-90]{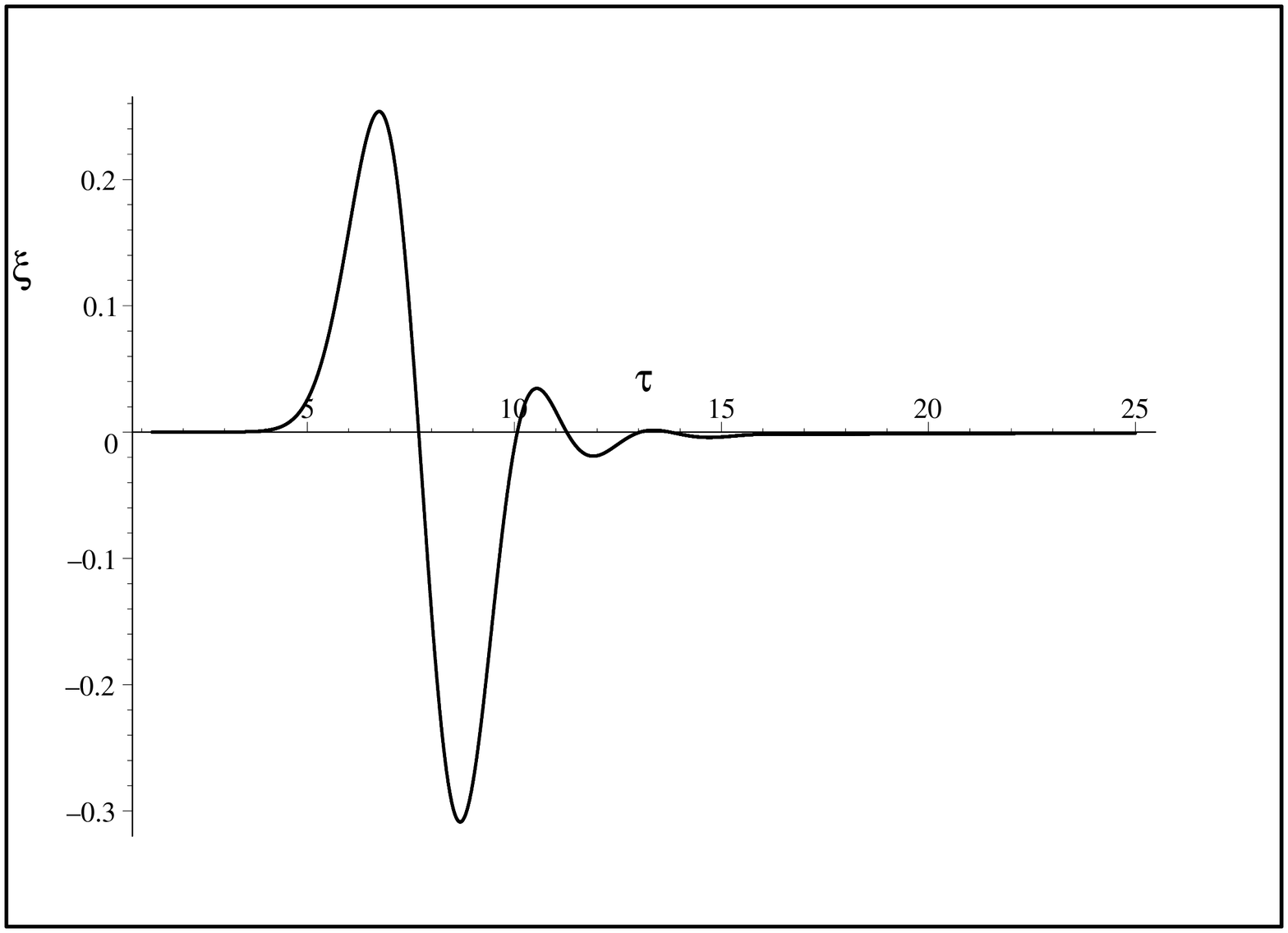}}
%\vspace{-1cm}
\caption{$\xi(\tau)$ as a function of $\tau$ for initial data of the form (\ref{eq09h}), for $R_0=0.25$ and $J=0.25$, that is $x=1.0$. Notice the strong damping of the oscillations after the arrival of the incoming pulse.}
\end{figure}

As a final example, we present the numerically computed evolution of $\xi(\tau)$ as a function of $\tau$ for the cases $R_0=0.25$ and $J=0.35$, that is $x=1.4$, given as the thin line curve in Fig. 7, and for $R_0=0.25$ and $J=0.55$, that is $x=2.2$, also given in Fig. 7 as the thick line curve. In these examples the QNR is over damped and no trace of ``ringing'' is apparent in the time dependence of $\xi(\tau)$. In particular, the last case, $x=2.2$, corresponds to $y= 4.84$ in the notation of \cite{nakao}, where in what they call the first class solutions one has a pure imaginary QNM frequency, and, therefore, we would have only a purely exponentially decreasing QNR. These plots confirm also the stability of the corresponding static configurations.
\begin{figure}
\centerline{\includegraphics[height=12cm,angle=-90]{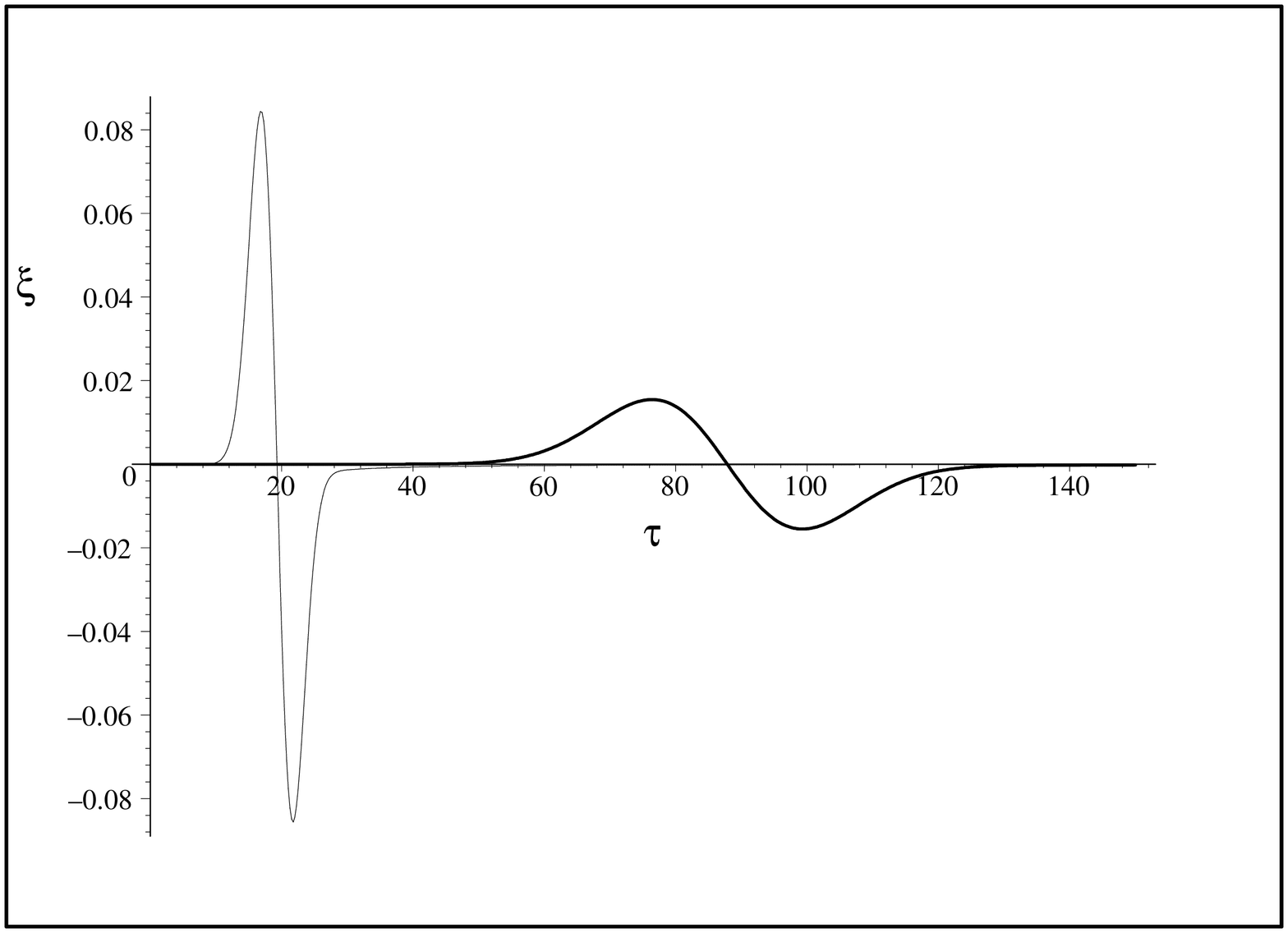}}
%\vspace{-1cm}
\caption{$\xi(\tau)$ as a function of $\tau$ for initial data of the form (\ref{eq09h}), for $R_0=0.25$ and $J=0.35$, that is $x=1.4$ (thin line), and  $R_0=0.25$ and $J=0.55$, that is $x=2.2$ (thick line). We see that the quasi normal oscillations are absent or completely damped out, and we only have the response of the shell to the incoming pulse.}
\end{figure}

\section{A note on the Price-Husain QNM toy model}

In this Section we consider the problem of the apparent discrepancy between the evolutions obtained using the formalism developed in \cite{gleram} and those described here. We shall illustrate the reasons for this by considering the QNM toy model given by Price and Husain in \cite{pricehus}, since, as will be seen, it bear a certain formal resemblance to the shell model discussed here. The Price and Husain model consists of two torsion strings, the first attached at $x=0$ and extending to $x=a$, where it is attached to a second string that extends to $x= \infty$. The strings have different characteristic constants, so that the respective equations of motion are,
\begin{equation}\label{eeq01a}
-\frac{\partial^2 u_1}{\partial t^2} +\frac{\partial^2 u_1}{\partial x^2}  = 0
\end{equation}
and,
\begin{equation}\label{eeq01b}
-k^2 \frac{\partial^2 u_2}{\partial t^2} +\frac{\partial^2 u_2}{\partial x^2}  =  0
\end{equation}
where $k>1$ is a constant. $u_1=u_1(t,x)$, and $u_2=u_2(t,x)$, represent the torsion angle respectively in the intervals $0\leq x\leq a$, and $a \leq x \leq \infty$. The boundary and matching conditions are,
\begin{eqnarray}\label{eeq02}
u_1(t,0) & = & 0 \nonumber \\
u_1(t,a) & = & u_2(t,a)   \\
\left.\frac{\partial u_1}{\partial x}\right|_{x=a}   & = & \left.\frac{\partial u_2}{\partial x}\right|_{x=a}  \nonumber
\end{eqnarray}

The general solutions of (\ref{eeq01a}) and (\ref{eeq01b}), are,
\begin{equation}\label{eeq03}
 u_1(t,x) = f(t+x) - f(t-x)
\end{equation}
where $f(y)$ is, in principle, an arbitrary function of $y$, and we have included the boundary condition at $x=0$, and,
\begin{equation}\label{eeq04}
u_2(t,x) = F(t+kx) +G(t-kx)
\end{equation}
where $F(y)$, and $G(y)$ are also functions of $y$.

One can now check that, imposing the matching conditions, and assuming that all the functions vanish for large values of their arguments, the general solution functions $f(y)$, $F(y)$, and $G(y)$ satisfy the relations,
\begin{eqnarray}\label{eeq08}
F(t+k x) & = &  \frac{1}{2 k} \left[ (k+1) f(t+kx-ka+a)-(k-1)f(t+kx-ka-a) \right] \nonumber \\
G(t-k x) & = &  \frac{1}{2 k} \left[ (k-1) f(t-kx+ka+a)-(k+1)f(t-kx+ka-a) \right]
\end{eqnarray}
and, therefore, an arbitrary $f(y)$ determines a complete solution of the problem. In particular, if we choose $f(y)$ of compact support, we will also get $F$ and $G$ of compact support. This type of solution describes then an incoming pulse of compact support that approaches the junction at $x=a$, excites a non vanishing $u_1$ for a finite time, and gives rise to an out going pulse, also of compact support. There is, in general, no ``ringing'', as no quasi normal mode is excited. This is precisely the situation considered in \cite{gleram}, where we could fix the function $\xi(\tau)$ arbitrarily, and, by choosing it of compact support we got also $\psi_-$ and $\psi_+$ of compact support, with no quasi normal ringing.

But the Price and Husain model does display QNR. To see how this comes about we assume that the functions admit Fourier transforms of the form,
\begin{eqnarray}
\label{eeq09}
f(y) &=& \int_{-\infty}^{\infty} e^{i \omega y} \tilde{f}(\omega) d \omega  \nonumber \\
F(y) &=& \int_{-\infty}^{\infty} e^{i \omega y} \tilde{F}(\omega) d \omega    \\
G(y) &=& \int_{-\infty}^{\infty} e^{i \omega y} \tilde{G}(\omega) d \omega  \nonumber \\
\end{eqnarray}
Replacing in the system (\ref{eeq08}) we eventually find that we may write,
\begin{eqnarray}
\label{eeq11}
f ( y) & = & 2\,\int_{-\infty}^{\infty} e^{i \omega y}{\frac {{{\rm e}^{ia\omega \left( 1+k
 \right) }}\tilde{F} ( \omega) k}{{{\rm e}^{2\,i\omega a}} \left( 1+k
 \right) -k+1}} d \omega \nonumber \\
G(y) & = & \int _{-\infty}^{\infty} e^{i \omega y}{\frac{ \tilde{F} ( \omega ) {{\rm e}^
{2\,i\omega ka}} \left[ {{\rm e}^{2\,i\omega a}} \left( -1+k \right) -1-k \right]
}{{{\rm e}^{2\,i\omega a}} \left( 1+k \right) -k+1}} d \omega
\end{eqnarray}

These expressions provide the solution for the problem of finding the evolution assuming that we are given an incoming pulse $F(t+kx)$ as the data for the problem, just as in the present shell problem. If $\tilde{F}(\omega)$ has suitable analytic properties it may be possible to compute the integrals in (\ref{eeq11}) by extending the path of integration  to the complex $\omega$ plane. We notice here that the denominators in (\ref{eeq11}) have simple poles at,
\begin{equation}\label{eeq12}
 \omega = \omega_n = n\frac{\pi}{a} + \frac{i}{2a} \ln\left(\frac{k+1}{k-1}\right)
\end{equation}
where $n$ is an integer, including zero. Notice that all $\omega_n$ have the same positive imaginary part, and, therefore, when it is possible to close the path of integration by adding an infinite semicircle in the upper half plane of $\omega$, each one of these poles will contribute a factor of the form $\exp(-\ln((k+1)/(k-1)) y/(2a))$ times an oscillating factor depending on $n$ and on the explicit form of $\tilde{F}(\omega)$. These are the characteristic features of the ``ring down'' associated with the quasinormal modes. In fact, if we go back to (\ref{eeq08}), and write it in the form,
\begin{eqnarray}\label{eeq13}
F(t) & = &  \frac{1}{2 k} \left[ (k+1) f(t-ka+a)-(k-1)f(t-ka-a) \right] \nonumber \\
G(t) & = &  \frac{1}{2 k} \left[ (k-1) f(t+ka+a)-(k+1)f(t+ka-a) \right]
\end{eqnarray}
and assume $f(y)=\exp(i \omega y)$ we get,
\begin{eqnarray}\label{eeq14}
F(t) & = &  \frac{1}{2 k} \left[ (k+1) \exp(i \omega a)-(k-1) \exp(-i \omega a) \right] \exp(i \omega (t-k a)) \nonumber \\
G(t) & = &  \frac{1}{2 k} \left[ (k-1) \exp(i \omega a)-(k+1) \exp(-i \omega a) \right] \exp(i \omega (t+k a))
\end{eqnarray}
The condition $F(t)=0$ corresponds to a purely out going wave for $x>a$. This is achieved precisely for $\omega$ of the form (\ref{eeq12}). Taken literally, these type of solution correspond to both $f(t-x)$ and $f(t+x)$ that decrease exponentially for $t \to +\infty$, but increase exponentially as $t \to -\infty$. Similarly, we have $G(t-kx)$ that grows exponentially for large $x$. These solutions are therefore unphysical. What is then the relation between these solutions and the poles in (\ref{eeq11})? The crucial point is that the poles are effective only if the integration path can be closed on the upper half plane of $\omega$. In general, for $k> 1$, this requires $ y +2 ka > 0$, and therefore, the exponential terms are not present for $y$, and as a consequence $t$, less than a certain lower bound.

But, why do we see a ring down in one case and not in the other? To understand what is happening here we go back to (\ref{eeq13}), and assume that $f(y)$ has a Fourier transform $\tilde{f}(\omega)$. Then, replacing in (\ref{eeq08}), we find,
\begin{equation}\label{33}
\tilde{F} ( \omega) = \frac{1}{2k}\, \tilde{f} (\omega) {
{\rm e}^{-i\omega ka}} \left[  \left( k+1 \right) {{\rm e}^{i\omega a}}-{{\rm e}^{
-i\omega a}} \left( k-1 \right)  \right]
\end{equation}
But this implies that for general $\tilde{f}(\omega)$, the transform $\tilde{F} ( \omega)$ will contain a factor that precisely cancels the denominators in (\ref{eeq11}), and therefore there will be no ring down in the solution of the problem, unless, of course, $\tilde{f} (\omega)$ itself contains the appropriate poles, as in (\ref{eeq11}). The general conclusion is then that both (\ref{eeq08}) and (\ref{eeq11}) provide a complete solution of the problem and are, therefore, completely equivalent. Clearly, for the purpose of making the QNR apparent it is simpler and, in a sense, more ``natural'' to use (\ref{eeq11}) and specify freely $F(y)$, i.e., the incoming pulse shape, just as was done for the shell problem in the present work.

\section{Final Comments}

In this paper we have considered again the perturbative analysis of the static configurations of the Apostolatos and Thorne shell model by modifying the formalism developed in \cite{gleram}. As a result we have been able to show the completeness of the mode expansion as regards the characteristic data problem for the perturbative dynamics of the shell. This, in turn, provides a simple and direct proof of the stability under bounded (symmetry preserving)  perturbations of the general static configurations of the shell. We have also derived a set of coupled linear ordinary and partial differential equations that describe the general perturbative evolution of the shell. This set of equation can be used to set up an initial value problem for the shell that can be solved numerically, but, more importantly, one can prove the existence of a positive definite constant of the motion, that implies the stability of the motion resulting for an arbitrary perturbation. In several examples considered we find perfect agreement between the full numerical evolution and that obtained through the numerical integration of the mode expansion, although they are completely different in detail. At first sight our results seem to be in contradiction with those of Kurita and Nakao \cite{nakao}. We believe that there is no contradiction here, and that our computations and those in \cite{nakao} are perfectly compatible. The problem appears because Kurita and Nakao conclude that the existence of certain complex zeros (or poles) in the complex $\omega$ plane automatically imply a direct effect on the evolution of the shell, but this is not necessarily the case. The usual arguments for relating these poles to the evolution relay rather heavily on the possibility of extending integrals on the real $\omega$ axis to the complex $\omega$ plane, in such a way that one effectively picks up the residues of those poles, but this may not always be possible, and, even, it might happen that one can disregard those poles, when they exist, by extending the integration path in the opposite direction. In fact, since the mode expansion involves Bessel functions in rather complex combinations, the matter of finding the extensions of appropriate function to the complex $\omega$ plane is a highly nontrivial undertaking. Going back to our own derivations, we notice that although we found QNM with a negative real part of $\omega$, we only find evidence of QNR for those with a positive real part for $\omega$. We remark once again that the mode expansion does not require explicit consideration of the complex $\omega$ plane, as only the real $\omega$ axis is involved. We believe that although the work \cite{nakao} is valuable as regards the finding of the QNM of the system, since these by themselves are unphysical (they diverge generally in some space or time direction), given the very complex nature of the functions and of their possible extensions to the complex plane, conclusions about the stability of the static configurations of the shell can only be reached by explicitly showing the relation between these modes and the general evolution of the shell. In fact, in view of our results concerning completeness of the mode expansion and its relation to stability, we can only conclude that all in principle unstable QNM must be suppressed, as well as possibly some of the stable ones. Of course, explicit confirmation of this conclusion would require performing and analyzing appropriate extensions into the complex plane, but that is completely outside the scope of the present research.

\section*{Acknowledgments}

This work was supported in part by CONICET (Argentina).


\begin{thebibliography}{99}
\bibitem{absomega} We must warn the reader that the use of $|\Omega|$ in the equations that follow applies only to expressions where $\Omega$ is real, and is just a shorthand for $\Omega$, if $\Omega > 0$, and for $-\Omega$, if $\Omega < 0$. This appears to be trivial for real $\Omega$, but is crucial for {\em complex} $\Omega$, because, as it should be clear, expressions such as (\ref{eq1d}), (\ref{eq4da}), or (\ref{eq5d}), with $|\Omega|$ in the argument of the Bessel functions, are {\em not} solutions of (\ref{ATeq2}) for complex $\Omega$. We must therefore replace appropriately $|\Omega|$ in any expression related to the solution of the evolution equations before we attempt to compute it for complex $\Omega$. This should also be taken into account in our discussion of quasi normal modes and in any other instance where we consider complex values of $\Omega$.
\end{thebibliography}
 \end{document}